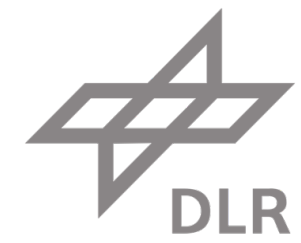

# An effect of abrupt current disruption

*Author:* *Andis Dembovskis,*
*Department of Satellite Systems,*
*Institute of Space Technologies,*
*German Aerospace Agency (DLR)*
*Bremen, Germany*
*Mail: andis.dembovskis@dlr.de*
*Phone: +49-421-24420-246*

ABSTRACT

Every engine, let it internal combustion engine in car or turbine of airplane, needs a high quality fuel igniter. During last decades there have been some minor changes made in ignition systems, like invention of Capacitive Discharge Ignition, Multiple Discharge Ignition, Ignition with Direct Current Discharge, but all based on the same principle of High Voltage spark path creation.

This work contains description, schematics and photographs of a new spark creation approach, providing high robustness through high power, big volume, long duration plasma. The system uses less or the same amount of energy as would CDI ignition, jet providing many times more efficient energy output. The solution is a highly applicable innovation, being able to significantly improve spark robustness in all current HV spark ignition systems.

Despite a simplicity of setup, it is still unclear why the effect persists, thus calling for additional research input.

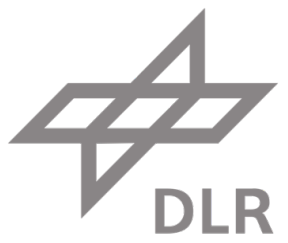

# 1 Introduction

There is observed an unordinary high voltage (HV) discharge effect. Using the same amount of stored energy in a capacitor, a rapid disturbance of HV current flow demonstrates very high intensity plasma discharge. Photographs taken with 6000fps camera confirm the observation.

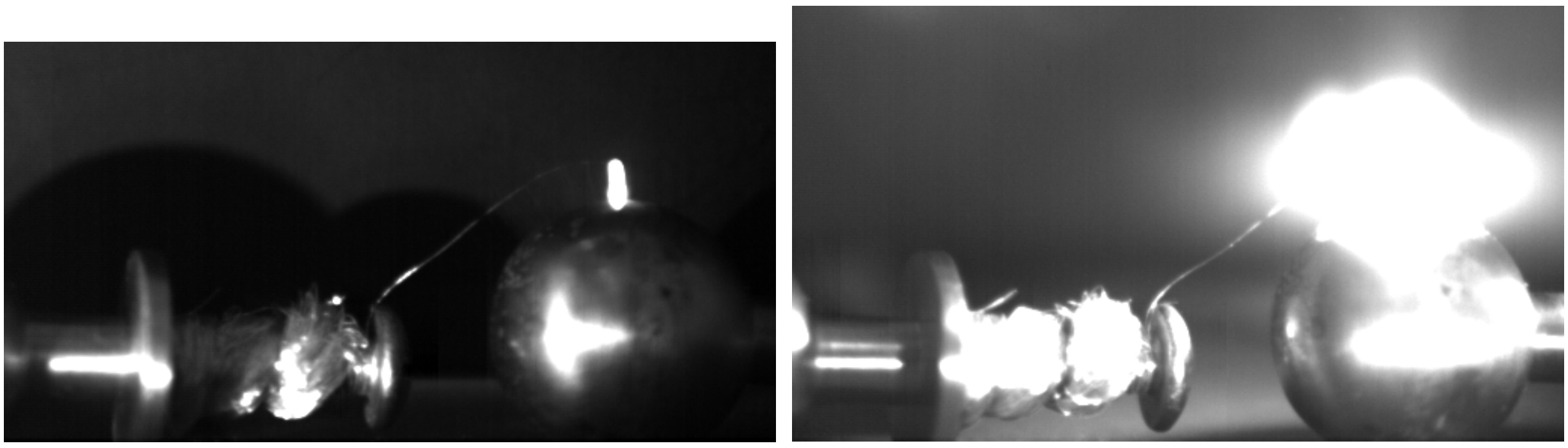

*Fig.1: Classic and disrupted HV current discharge*

Full list of photography sequences: http://www.andis.me/pub/plasma_photos.zip

The observed enhancements are: bigger plasma volume, acoustic shock wave, kinetic energy of the string electrode.

# 2 Schematics

Here below are two different simple schematics of capacitor discharge through CDI Coil (Capacitive discharge ignition coil, a part of ignition systems in cars).

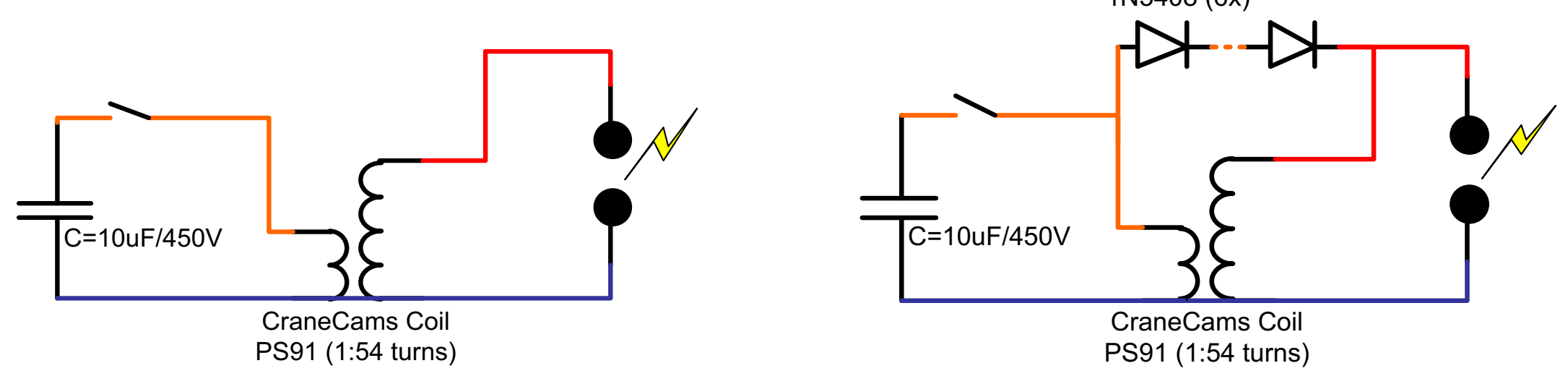

*Fig.2: Capacitor discharge. (a)On the left, in classic way. (b)On the right, with additional diodes.*

Although the addition of diodes in the second circuit would not be awaited to influence intensity of the spark discharge, the practical difference is huge and obvious. The additional diode-connection will be further referred as BD (Booster Diodes).

The principal idea is taken from [1], which is follow-up research of anomalies described in FireStorm spark plug research and replication activities [5],[7], as well as Stanley Meyer's unconventional, highly efficient hydrogen generation applications [3],[6].

The next scheme includes also charging of capacitor, producing a discharge every 2 seconds.

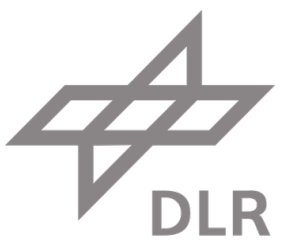

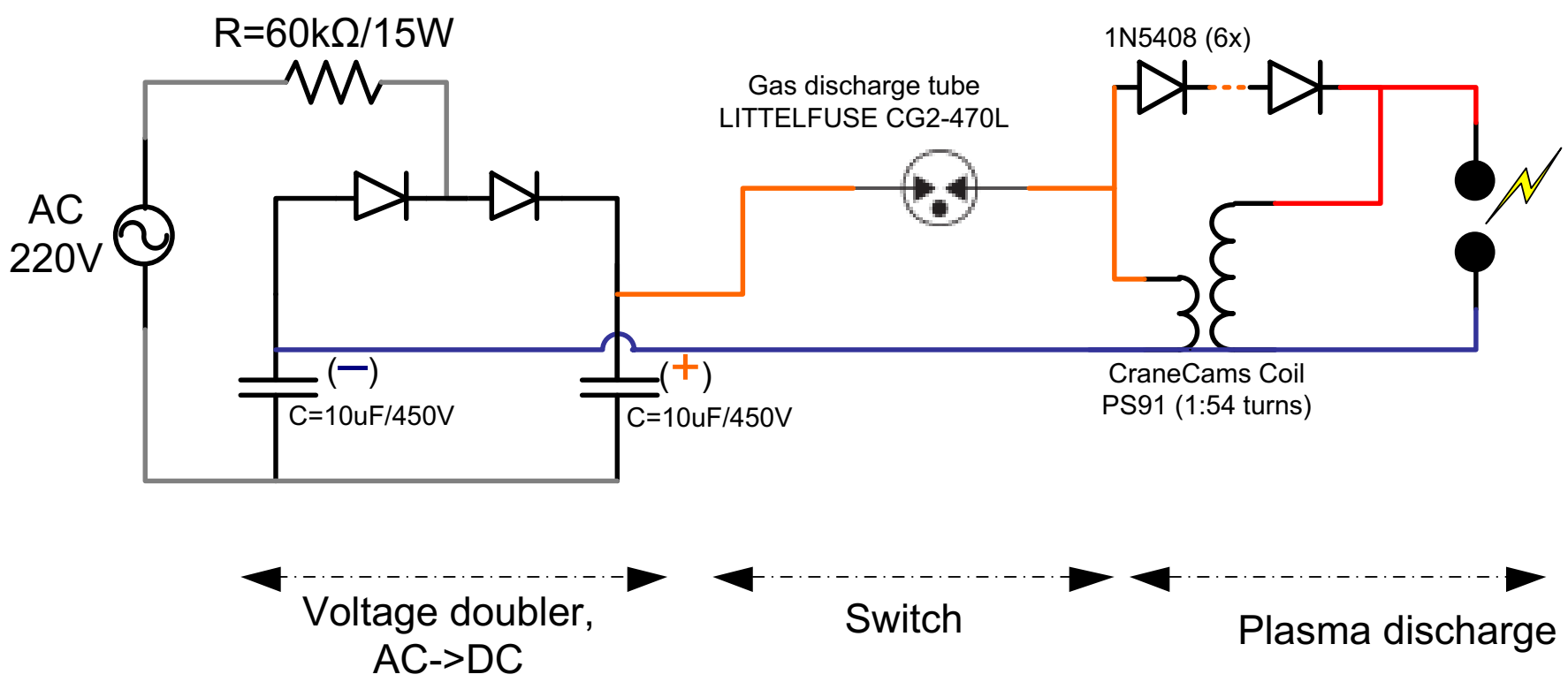

*Fig.3: Plasma discharge generator*

## 3 Current flow considerations

Case Fig.2a: the capacitor is discharged directly to the CDI coil. Assuming the cap being charged to 325V, the output of the tested coil had to be 325*54=17.5kV. A measurement with oscilloscope was not performed; however with eyes and photo camera the spark looks similar to the one in car spark plugs.

Case Fig.2b: the output HV from the coil for the first sees the ground through the attached diodes, since they are “open”, thus starts to flow that path. As soon as the current through the diodes reaches “surge current”, its flow is abruptly disrupted by rapid diode shut down. As the next easiest escape path is spark gap, plasma discharge is initiated. With eyes and photo camera there can be observed a big plasma ball around discharge electrodes, also visible that the wire gets kinetic energy – is swinging afterwards. In addition the sound power is higher – louder, well noticeable by hearing.

Case Fig.3: a charging circuit option is showcased. The setup works from 220V AC wall socket. It charges two caps, in configuration as voltage doublers. As soon as caps have reached 470V potential, the spark of gas discharge tube flashes over the fuse becoming a good conductor. At the moment caps are connected to coil – the discharge occurs, which generates plasma. Comparing to the 2b case, it is noticeable, that instead of 1uF of 325V, here supply power is 20uF 470V. But in this case the gas discharge tube itself consumes some energy from caps within flashover process. Still the big plasma ball effect is very obvious. The output energy is lowered in this case also because the caps don’t get full discharge, since the flashover within gas discharge tube stops somewhere between 90..0V.

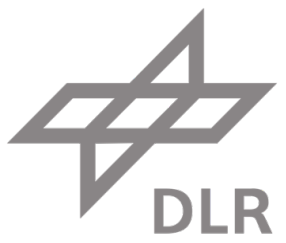

# 4 Measurements

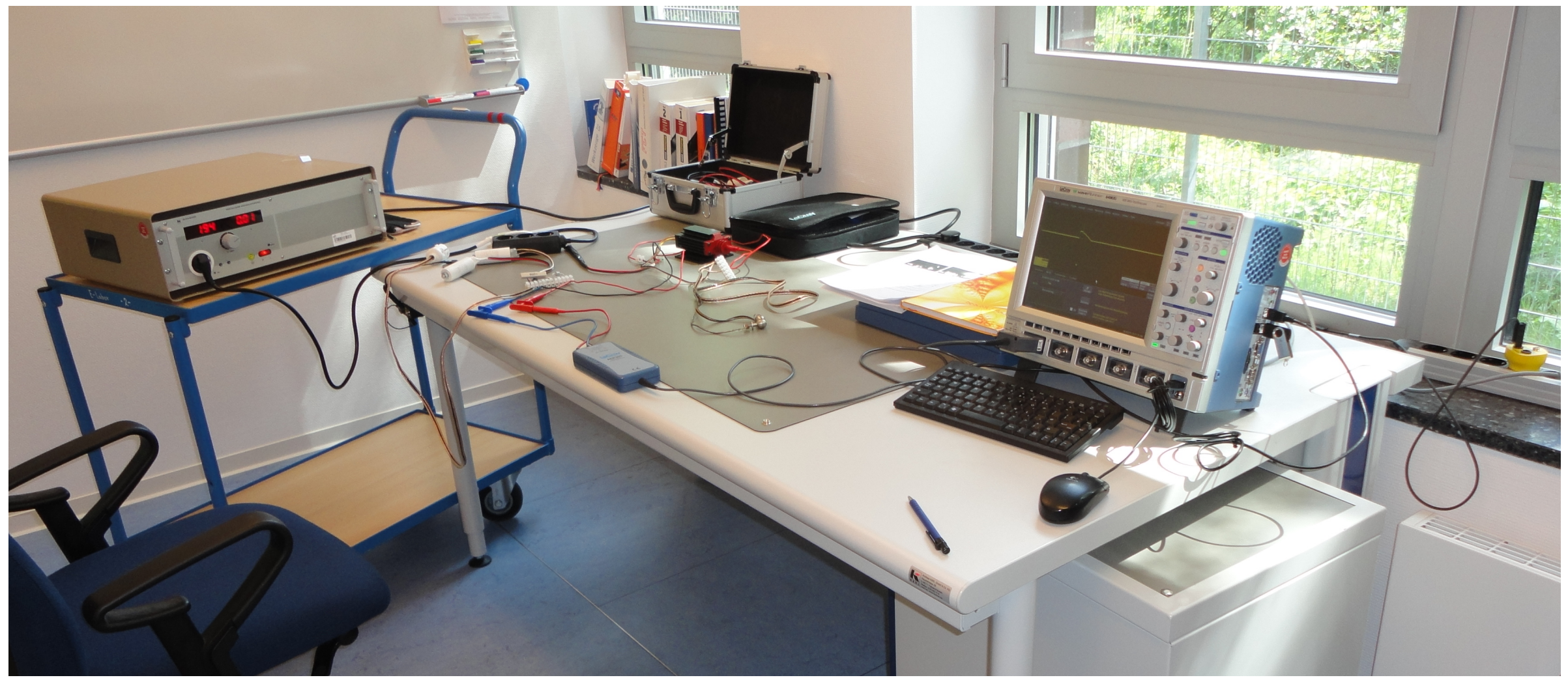

*Fig.4.1 Measurement facility*

## 4.1 Scope measurements for hand switch configuration

For simplification, the following per hand switched circuit was used within measurements:

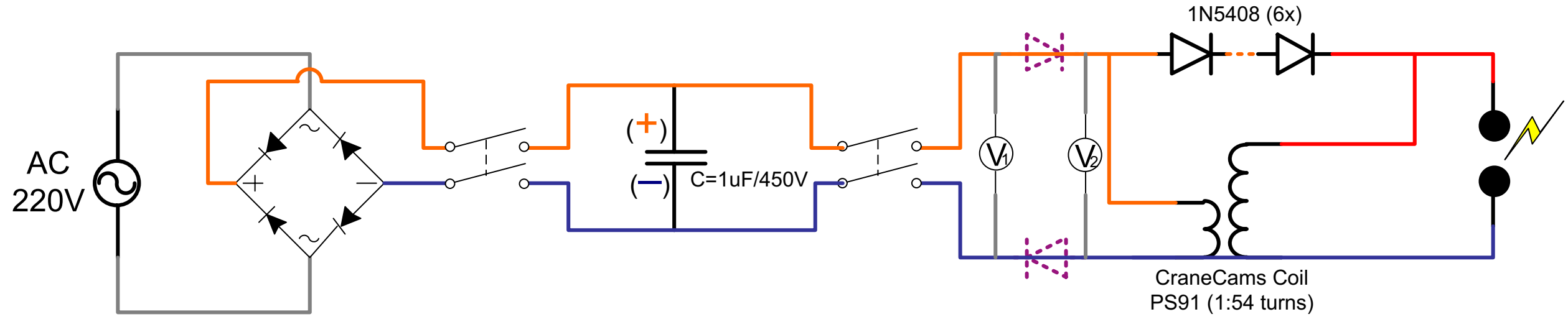

- Optional diodes, not changing the observed boost effect

- Voltmeter, LeCroy, 100MHz, 1400V peak max, 1000MΩ

*Fig.4: Circuit for Scope measurements*

With respect to the previous scheme in Fig.3, here is manual hand switching used. This allows using simple capacitor discharge, avoiding 50Hz socket influences. The additional diodes, noted above in violet color, were used to ensure unidirectional current flow. The boosting effect is similar with and without the optional diodes.

### 4.1.1 Spark power considerations

Consumed resources per spark:

- Charge capacity of the capacitor is Q=F*V=10e-6*325=0.00325 [Coulombs]
- Charging the 325V 1uF capacitor takes and stores $J=½*F*V^2=0.5*1e-6*325^2=0.05281$ [Joules]

To compare with, a 12V TCI (Transistor Controlled Ignition) circuit charging ignition coil takes 650us to charge with 20A, thus consuming:

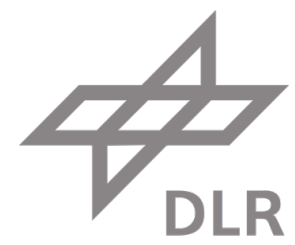

- Q=t*I=650e-6*20=0.013 [Coulombs]
- J=Q*V=0.02*12=0.156 [Joules]
- Assuming ignition is fired 10 times in a second, then the unit consumes W=Q*10*1=1.56 [Watts] in total.

### 4.1.2 Measurements

Below are measurement results of four cases: before and after capacitor diodes (V1 or V2), with and without booster diodes (BD).

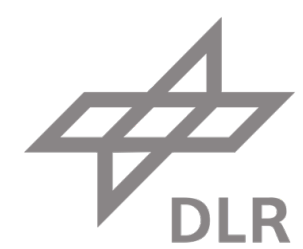

Oscilloscope measurements, 10us

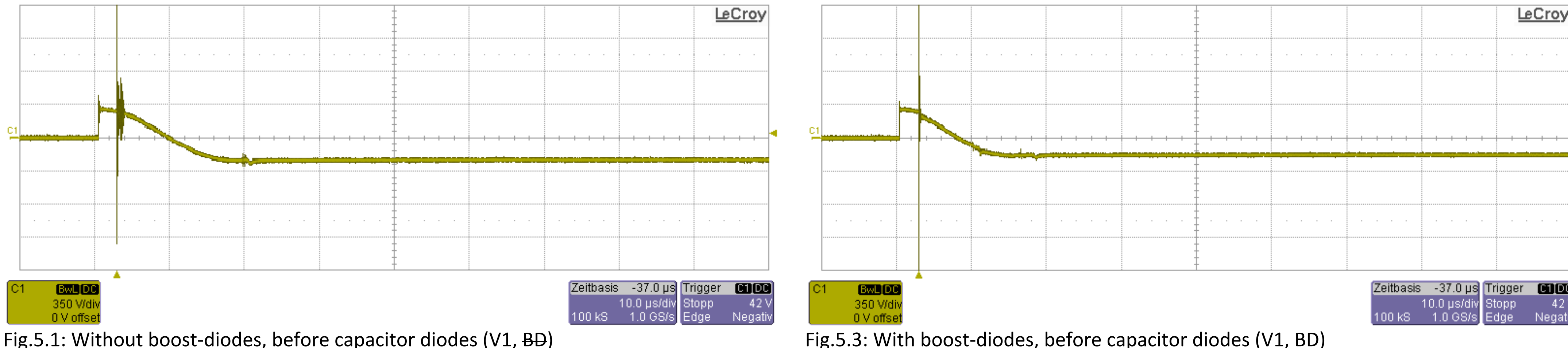

Fig.5.1: Without boost-diodes, before capacitor diodes (V1, ~~BD~~)

Fig.5.3: With boost-diodes, before capacitor diodes (V1, BD)

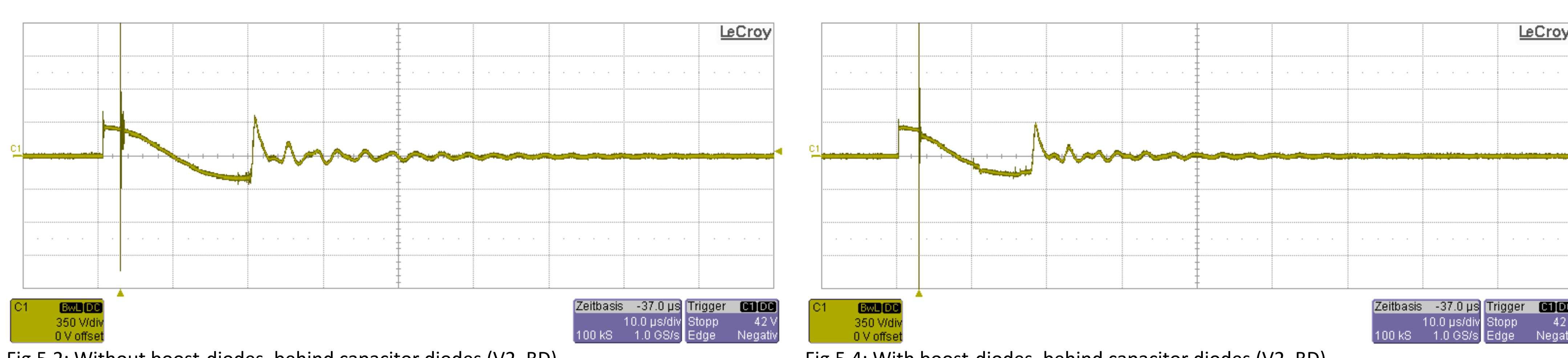

Fig.5.2: Without boost-diodes, behind capacitor diodes (V2, ~~BD~~)

Fig.5.4: With boost-diodes, behind capacitor diodes (V2, BD)

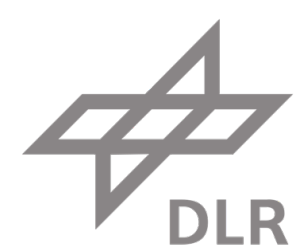

Oscilloscope measurements, 200ns, the high pulse from previous image

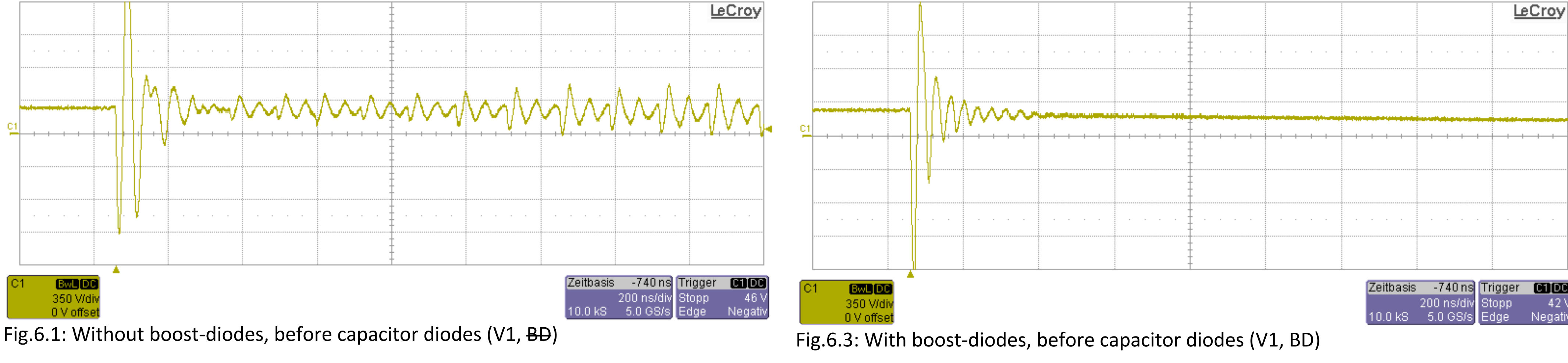

Fig.6.1: Without boost-diodes, before capacitor diodes (V1, ~~BD~~)

Fig.6.3: With boost-diodes, before capacitor diodes (V1, BD)

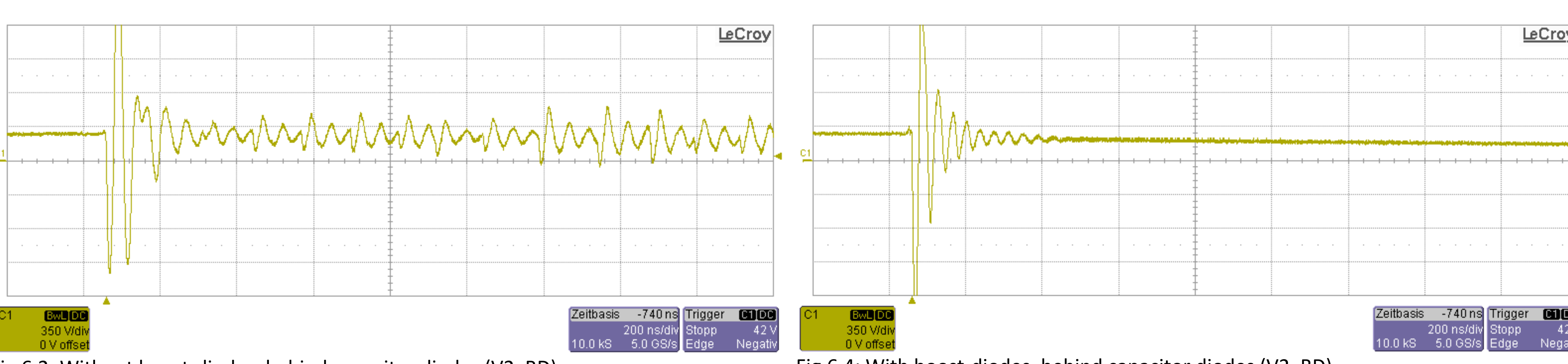

Fig.6.2: Without boost-diodes, behind capacitor diodes (V2, ~~BD~~)

Fig.6.4: With boost-diodes, behind capacitor diodes (V2, BD)

The following pictures show scope measurements for the case without directional-diodes (without the violet diodes within the Fig.4):

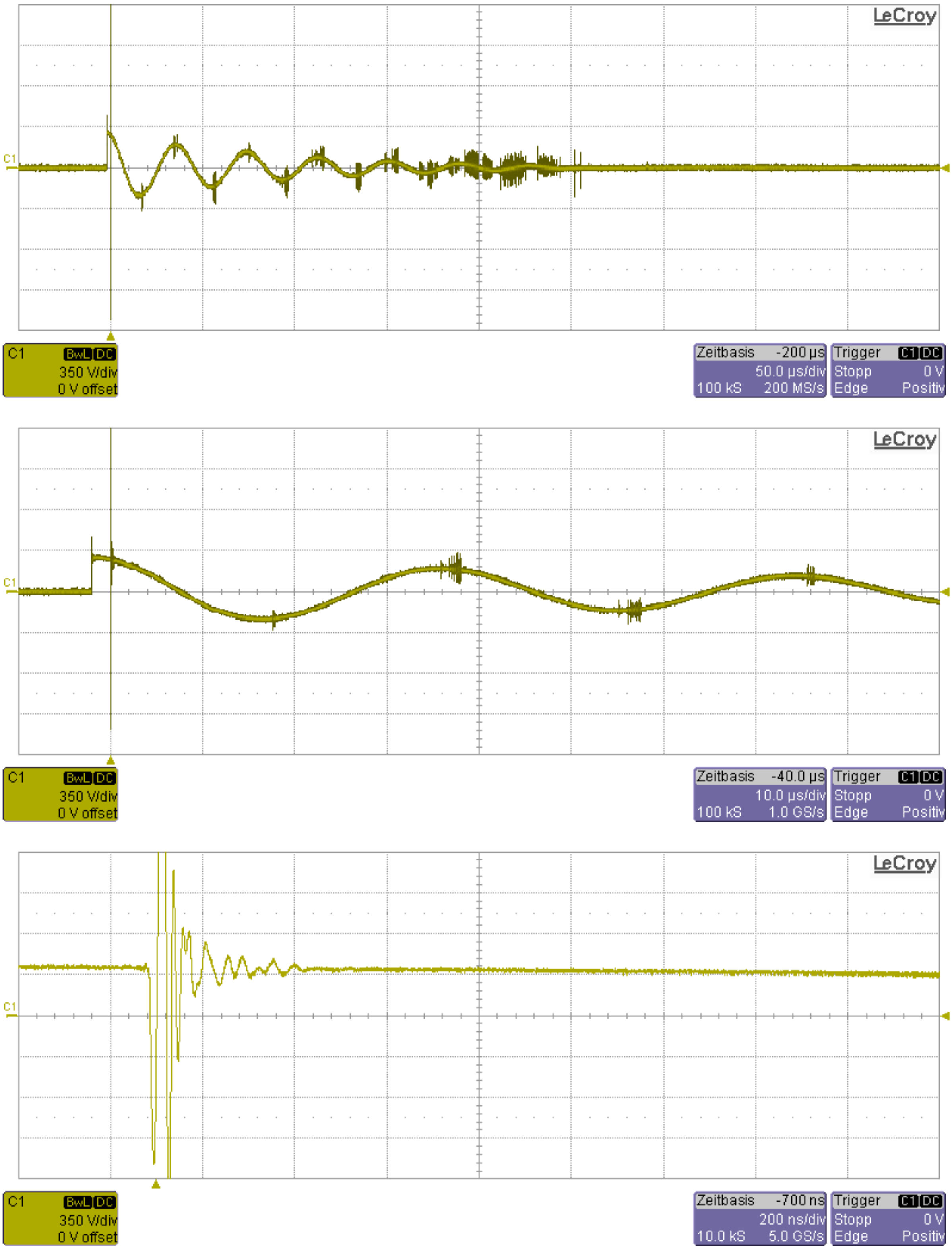

Fig.7: Scope measurements for case without directional diodes, ~~BD~~

And here the same scenario with booster diodes:

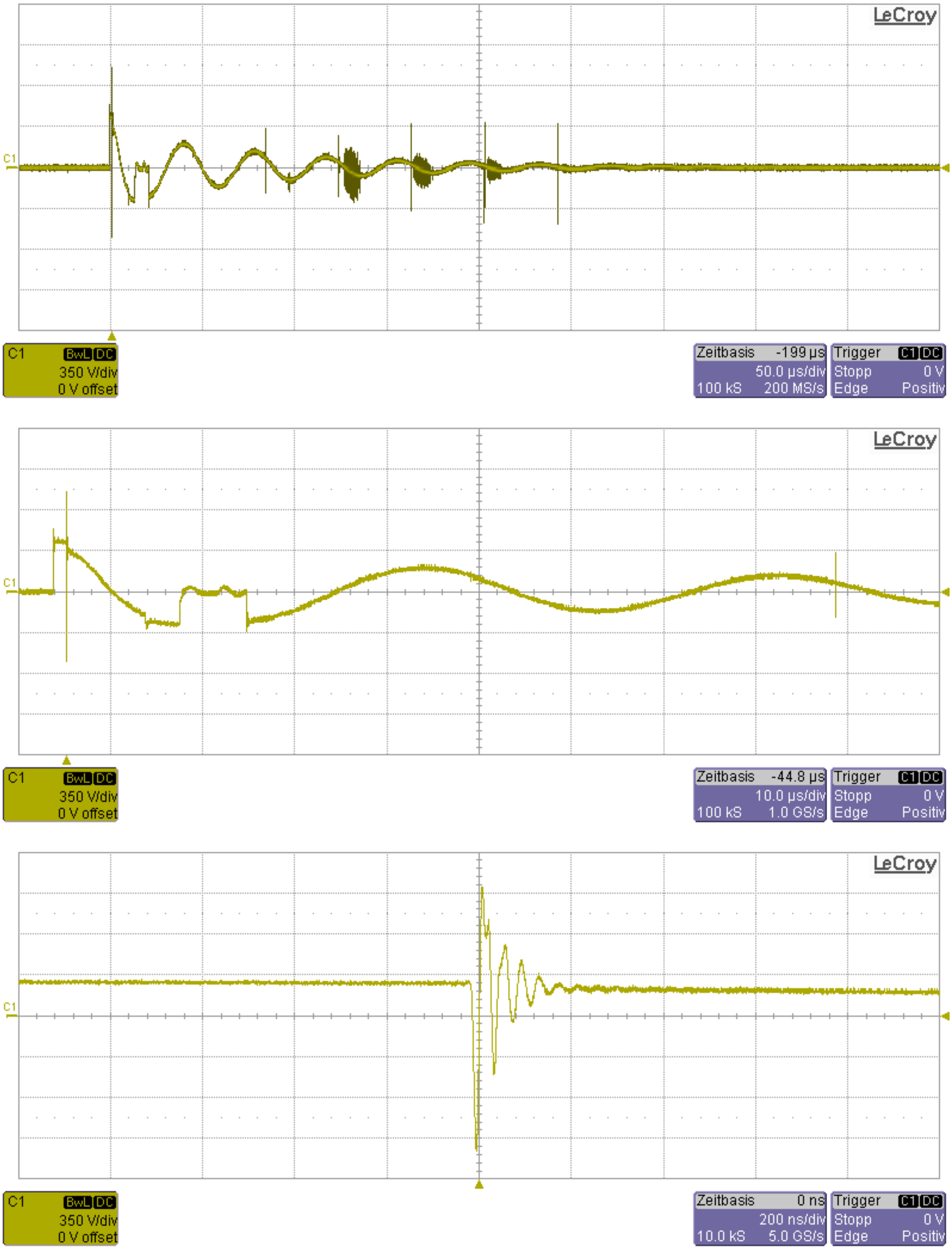

Fig.8: Scope measurements for case without directional diodes, with BD

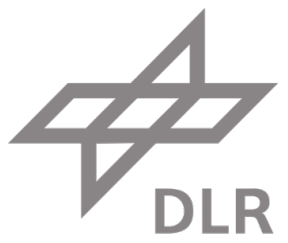

## 4.2 Hand switch scope measurement analysis

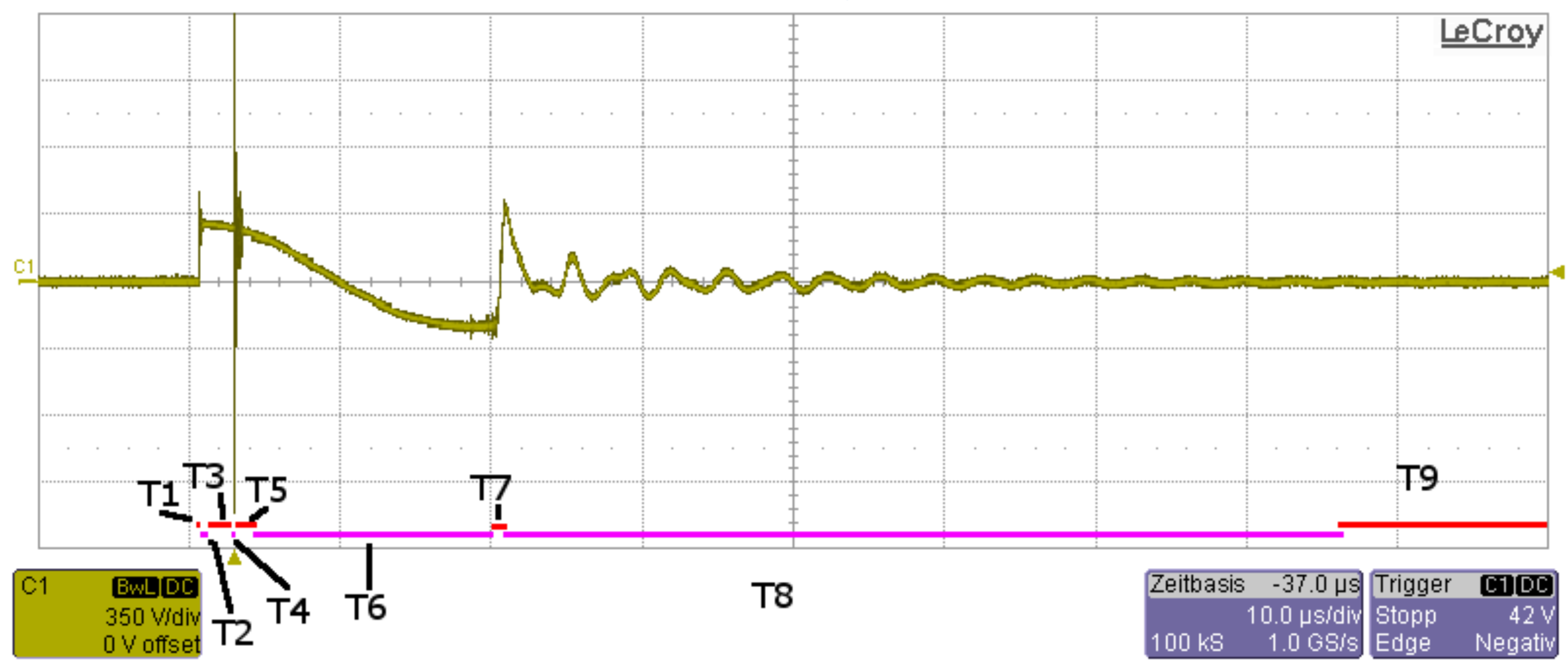

*Fig.9: Zones of analysis*

### 4.2.1 Analyses based on Fig.5, Fig.6

T1: Switch on time. The DPDT switch used within a setup is a mechanical switch, switching on permanently. Also within a zoomed version one sees straight growth.

T2: Oscillation time for the switch on voltage. It is an ordinary effect, awaited in any semiconductor circuit with rapid switching.

T3: Steady state time. Charging the coil primary winding. Voltage is slowly decreasing within this time.

T4: Primary winding response impulse. Because of rapid switch-on time. Its first pulse is negative, around -1200V for simple case and even more for case with booster diodes. (Due lack of HV measurement probes, more precise data are missing.) One can observe that the fist negative pulse is stronger with BD present however next swinging positive pulse seems to be higher for case without BD.

One can also speculate that at the time of sudden low voltage drop or within next oscillation, there is the HV pulse generated. And since in BD case the voltage drop is deeper, it influences the generated HV pulse to be higher.

T5: Oscillating time from the T4 impulse. In Fig.6.3 and Fig.6.4 it is obvious, that in the case of BD, the T4 pulse has a smooth dumped oscillation. Whereas without the BD, oscillations seem to be chaotic and also longer.

T6: Cap discharge. The first strange effect seems to be the fact that independent of one-way diodes behind the cap, it is not just discharging, but also charging with reverse voltage. This is to be explained with coil charge at the first half period and discharging later to the cap, thus acting like LC oscillating circuit of one cycle. One will note that T3+T6 forms like exponential curve. This also fits to classical EM theory of LC circuit. Another strange effect shows that in case with BD, the discharge cycle is restarting a little lower voltage than the curve of T3 would be expected to continue with sinusoidal shape. It shows like some extra energy was consumed from cap during the T4.

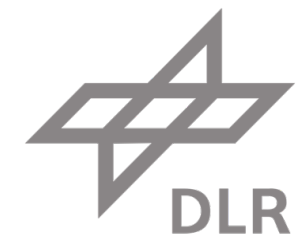

Additionally one will note that the cap discharge time is shorter for the BD case. If comparing T1 to T7 distance, in Fig.5.2 it is like 20us, whereas in Fig.5.4 it is 18us.

T7: A strange peak occurs. If comparing Fig.5.1 with Fig.5.2, one can note the small bounce of voltage in Fig.5.1 at the T7 time. This effect is more persistent at Fig.5.1 than in the Fig.5.3, relating also to showcase that the peak at Fig.5.2 is higher than at Fig.5.4. This might be diode shut down time, due the leakage current property.

One can a second time speculate to guess that the spark is happening at this point T7, thus indicating some potential changes induced from secondary to primary coil.

T8: (V2 only) Chaotic dumping oscillations of the T7 pulse.

T9: Steady state. One should note, that T9 at V1 case starts from T8 and takes non-zero state around -280V. This is to be explained that reverse current diodes do arrest the reverse charge within cap, not allowing it to flow opposite direction. Thus also one concludes the ability to hold up some bit of charge in the cap, not using it completely out for spark. Before next charge of the cap, for optimization purpose, one could reverse it, to decrease the consumed power for charging the cap.

Another observation: comparing the charge levels of the steady state T9 between Fig.5.1 and Fig.5.3, one notes that without BD it is ≈175V, whereas with BD is ≈230V. Both cases indicate that energy from the cap is consumed only within the phase T3 (broader speaking the T1…T4). Since the cap charge at the end, as Fig.5.1 and 5.3 indicates, is diminished only by that part.

A practical test showed, that from a single-charged 1uF cap (325V), it was feasible to make 4 discharges, by swapping connector cables after spark. Energy level (plasma ball size, brightness and bang volume) diminished with each discharge.

Conclusion1: There is not jet found straightforward supply voltage difference indicators to conclude why the booster diodes give plasma ball effect.

Conclusion2: With BD configuration, remaining reverse voltage in cap lower, indicating more consumed energy.

Conclusion3: With BD configuration, discharge + recharge time of the cap is for 2ms shorter.

Conclusion4: With BD configuration, the T7 peak is of lower energy.

Conclusion5: With BD configuration, the T5 contains harmonic dumped oscillations instead of chaotic.

Conclusion6: With BD configuration, the peak in T4 contains firstly stronger negative peak, and weaker positive.

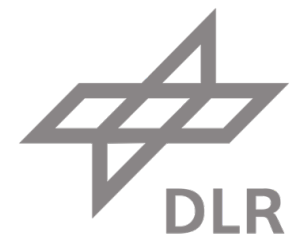

### 4.2.2 Analysis based on Fig.7, Fig.8

For the case without directional diodes, the measurement differs within the period T4 and significantly differs after T6.

For the T4: it can be observed that for the case without BD, the peaks are of significantly higher intensity. With BD they oscillate within range of 1400V.

For the T5, again, for non-BD case there are chaotic dumped oscillations and with-BD case, harmonic dumped oscillations.

For the T7, one observes missing pulse like in Fig.5.2 and Fig.5.4. Instead, for non-BD case at T6, T7, T8 it is smoothly dumped oscillations. And for with-BD case, there is a rapid zero-tooth, 10us long, within a middle of running dumped oscillations.

This allow again to speculate, that within T7 there is occurring plasma spark, since at that time circuit is short-connected over spark gap.

From measurements, one can estimate, that the capacitor with primary coil makes oscillations at period of 37.7us (≈26.5kHz), from Fig.8.1, and the oscillations at the T7 are at 38.6ns (25.8MHz), from Fig.8.3 alike. Thus approximately 1000 times higher.

The LC oscillation frequency of cap and primary coil, derived from specification of the used hardware:

```
L = 5.5e-3;
C = 1e-6;
w=1/sqrt(L*C)    = 13484
f = w/(2*pi)     = 2.146e+3 [Hz]
```

whereas if calculated the L from measurements taken:

```
T2 = 37.7e-6;
f2 = 1/T2         = 26.525e+3 [Hz]
w2 = f2*2*pi      = 1.6666e+5
L2 = 1/(w2^2*C) = 360.02e-3  [H]
```

It is very probable, that inductivity specification from the primary winding in the used technical specification file is mistaken [Ref.4].

Conclusion1: There is an obvious difference between BD and non-BD case comparing Fig.8.2 and Fig.7.2 – the zero-tooth in BD case within T7, 10us long.

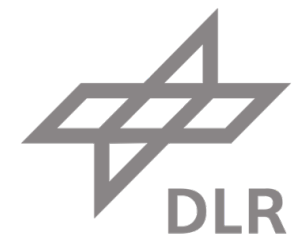

## 4.3 Scope measurements for plasma switch configuration

For another perspective view to measured values, here some additional details. These measurements were taken for the scheme of Fig.3, before and after the plasma switch.

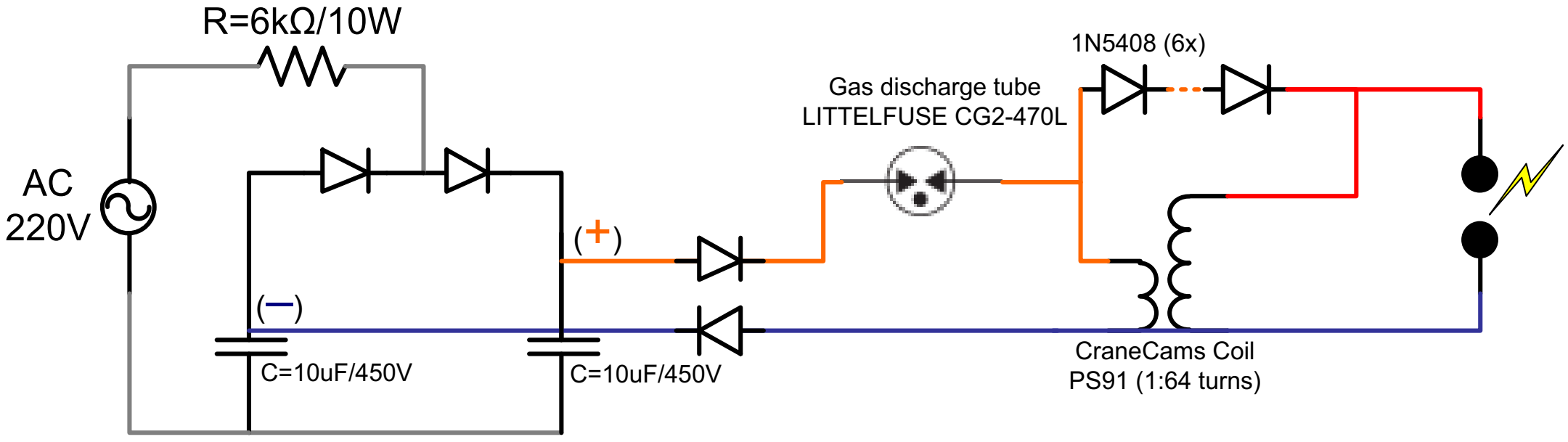

*Fig.10: Scope measurement positions for plasma spark generator*

As one notes, here is no the directional diodes present. This is unnecessary, since the gas discharge tube switch gets open as soon as voltage reaches zero. Thus there also is no present negative capacitor charge and no oscillations at V1.

The picture below shows switching speed of the gas tube.

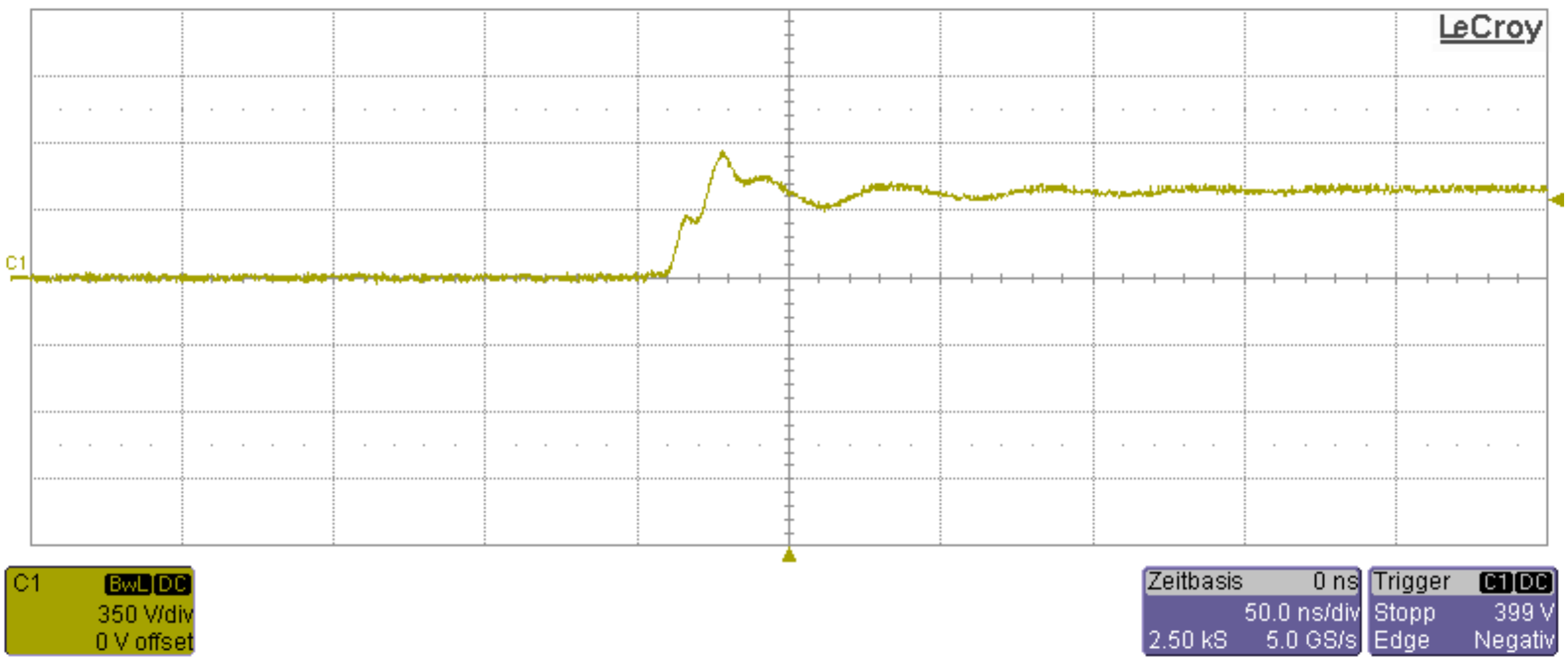

*Fig.11: Switch on time, V2, BD*

The picture above shows that switch on occurs within 20 ns. Although plasma interaction is expected to make some non-sharp switch effect, this still questions precision of the probe of the scope. The used HV scope probe is intended for precision up to 100MHz, meaning 10ns, thus lying at the border of resolution precision. Independent of that, one can still observe and hold as valid the small oscillations directly after switch on time. And also account that some IGBT switching transistor like AUIRGP50B60PD1, with Voltage_Rise_Time 10..20ns and Voltage_Turn_On_Delay_Time 20..40ns. This could substitute the plasma switch part with low voltage controlled switching.

Oscilloscope measurements, 10us

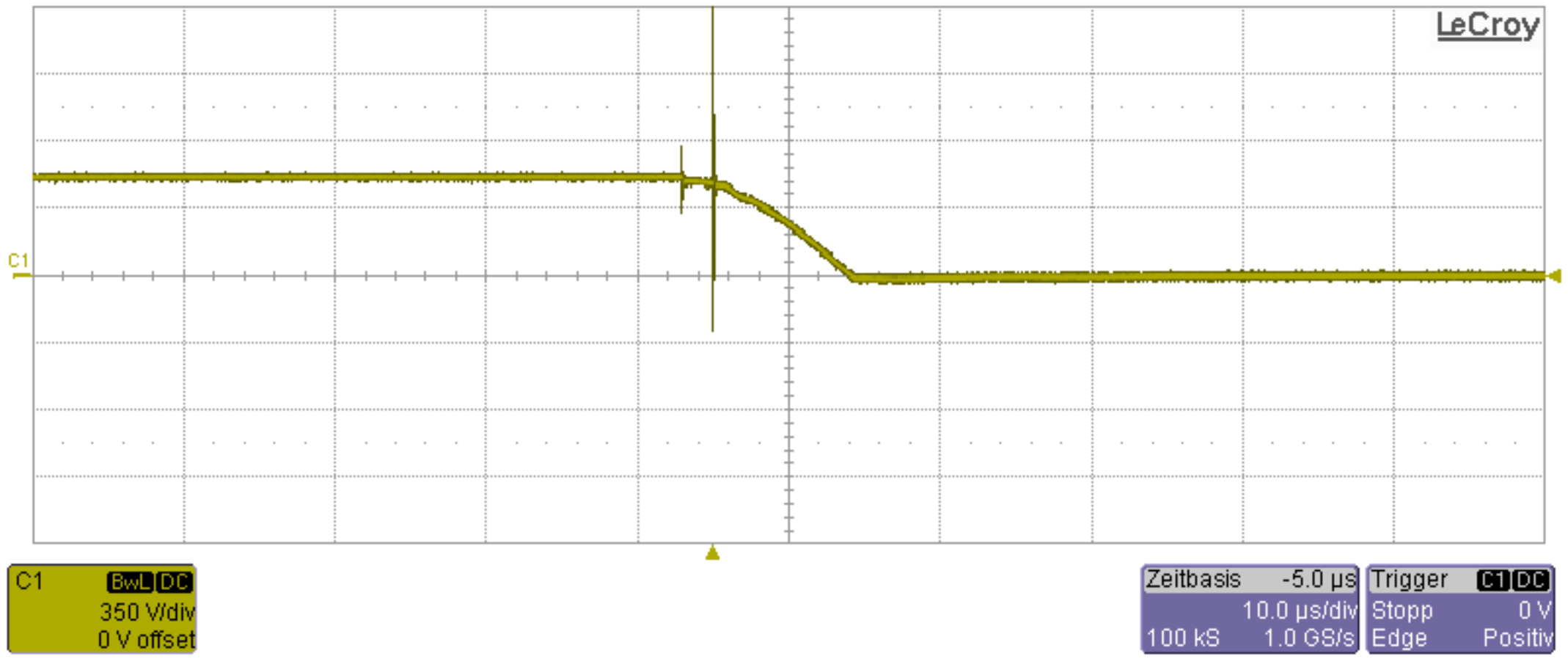

Fig.12.1: V1, ~~BD~~

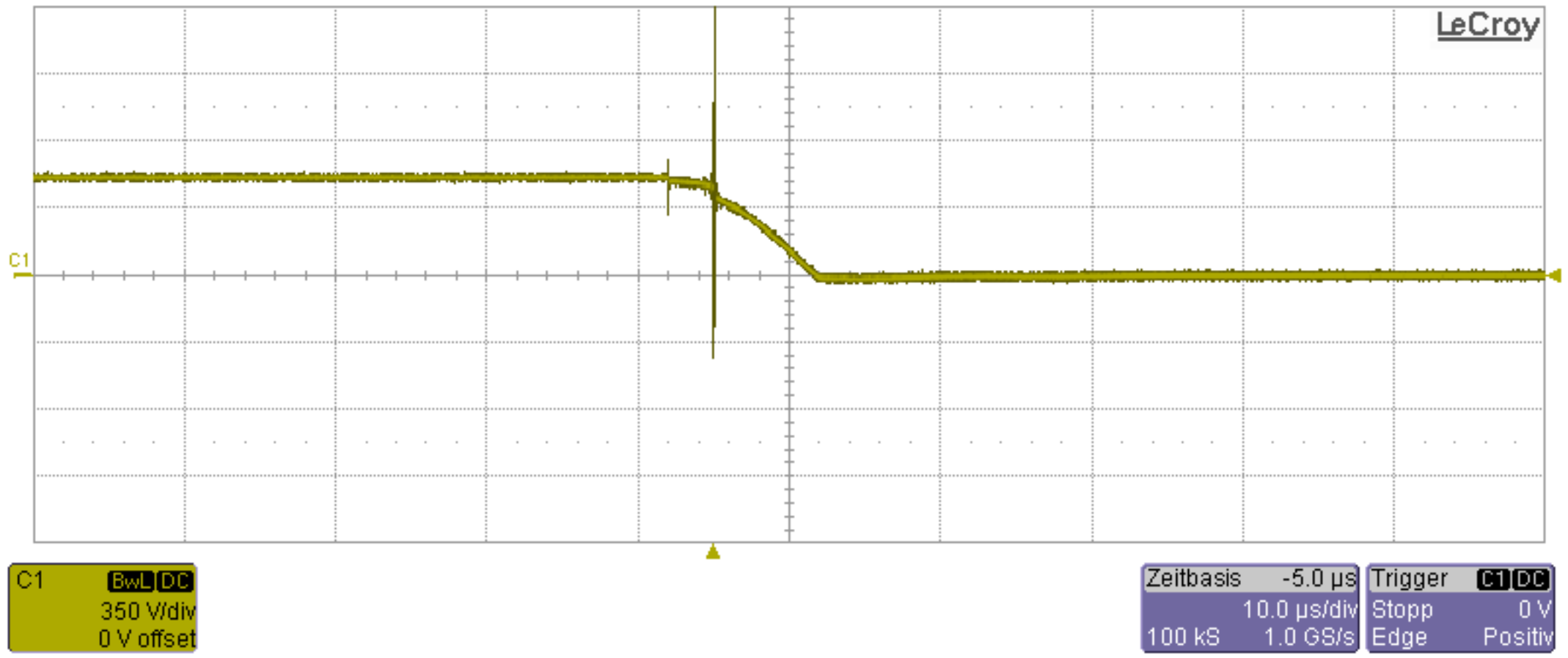

Fig.12.3: V1, BD

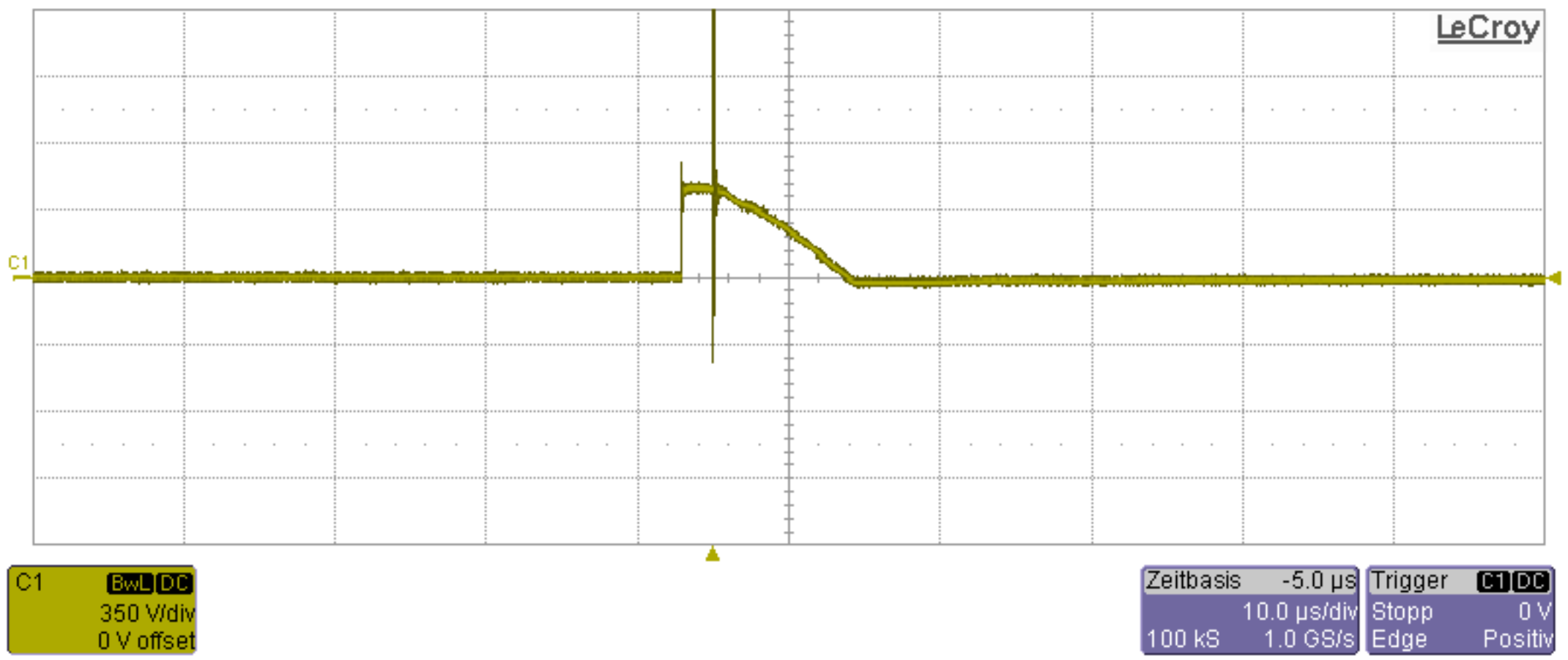

Fig.12.2: V2, ~~BD~~

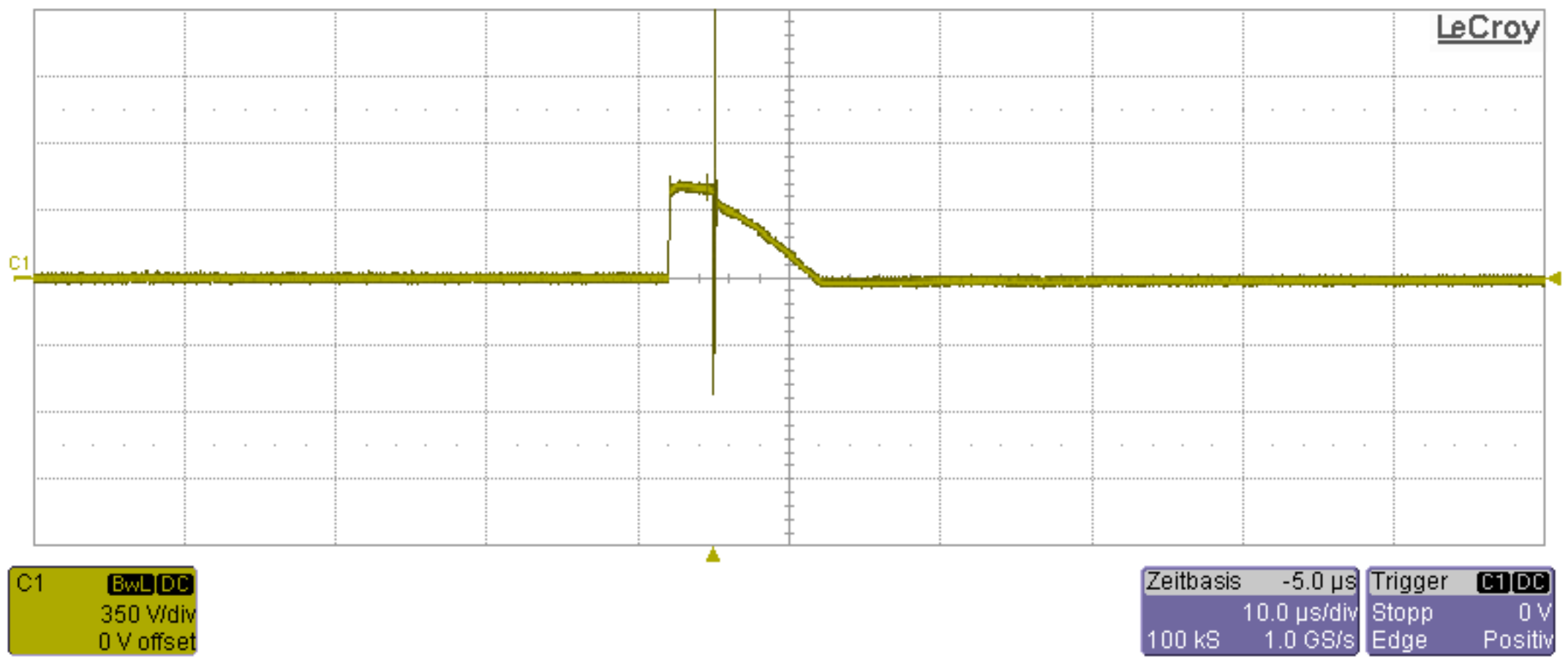

Fig.12.4: V2, BD

Oscilloscope measurements, 1us

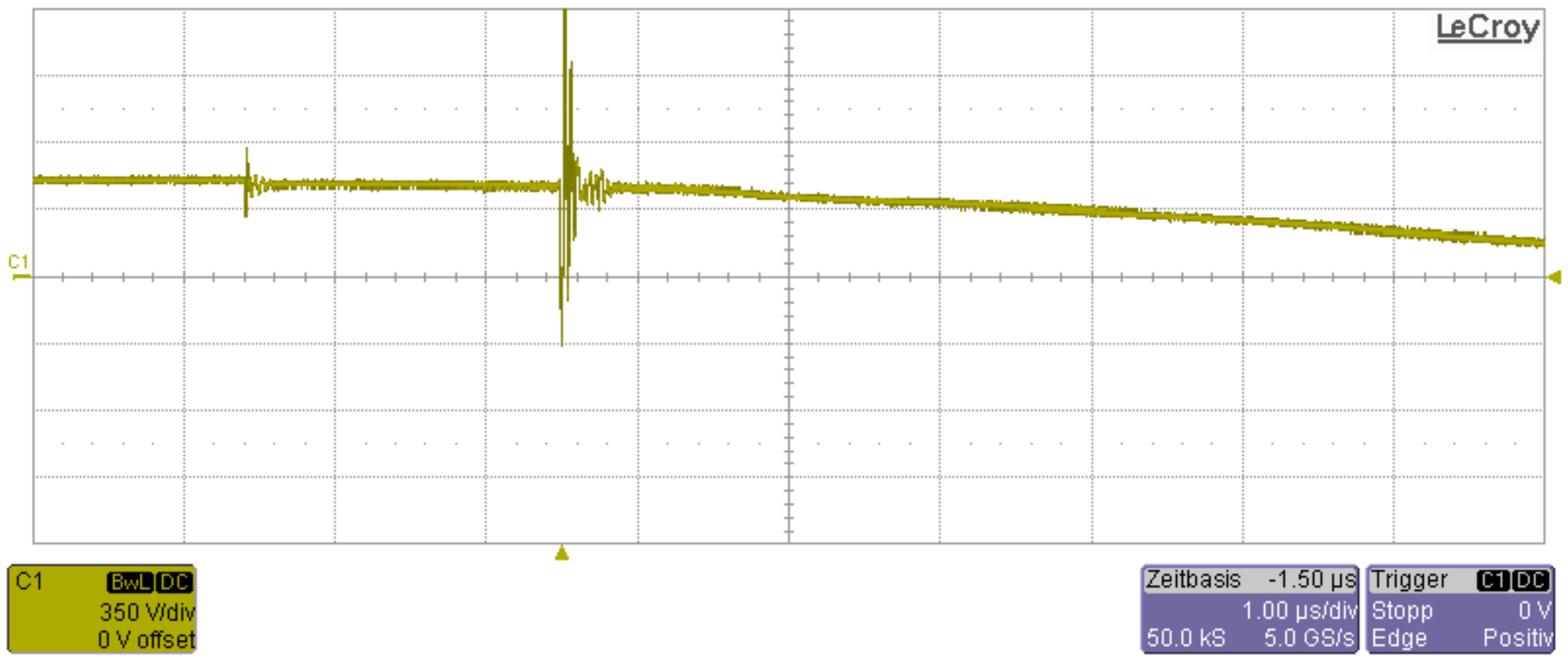

Fig.13.1: V1, ~~BD~~

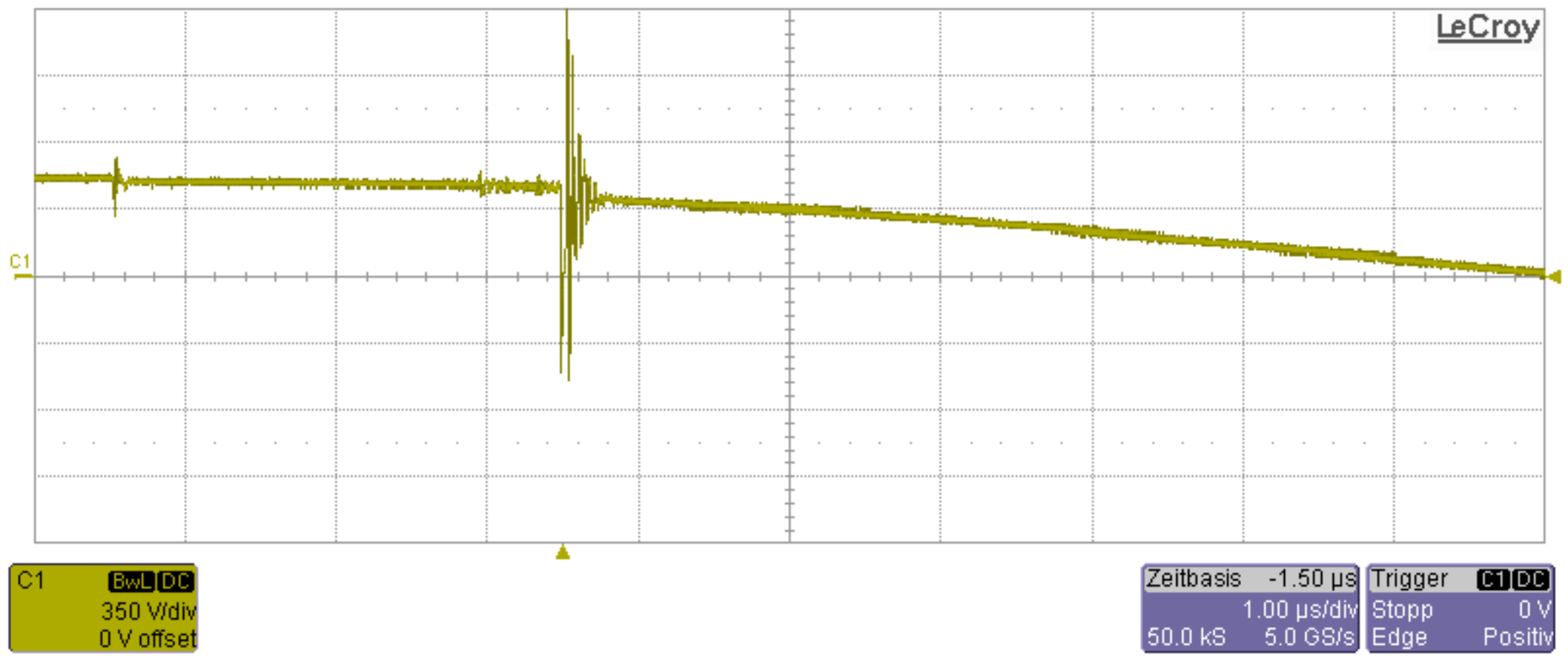

Fig.13.3: V1, BD

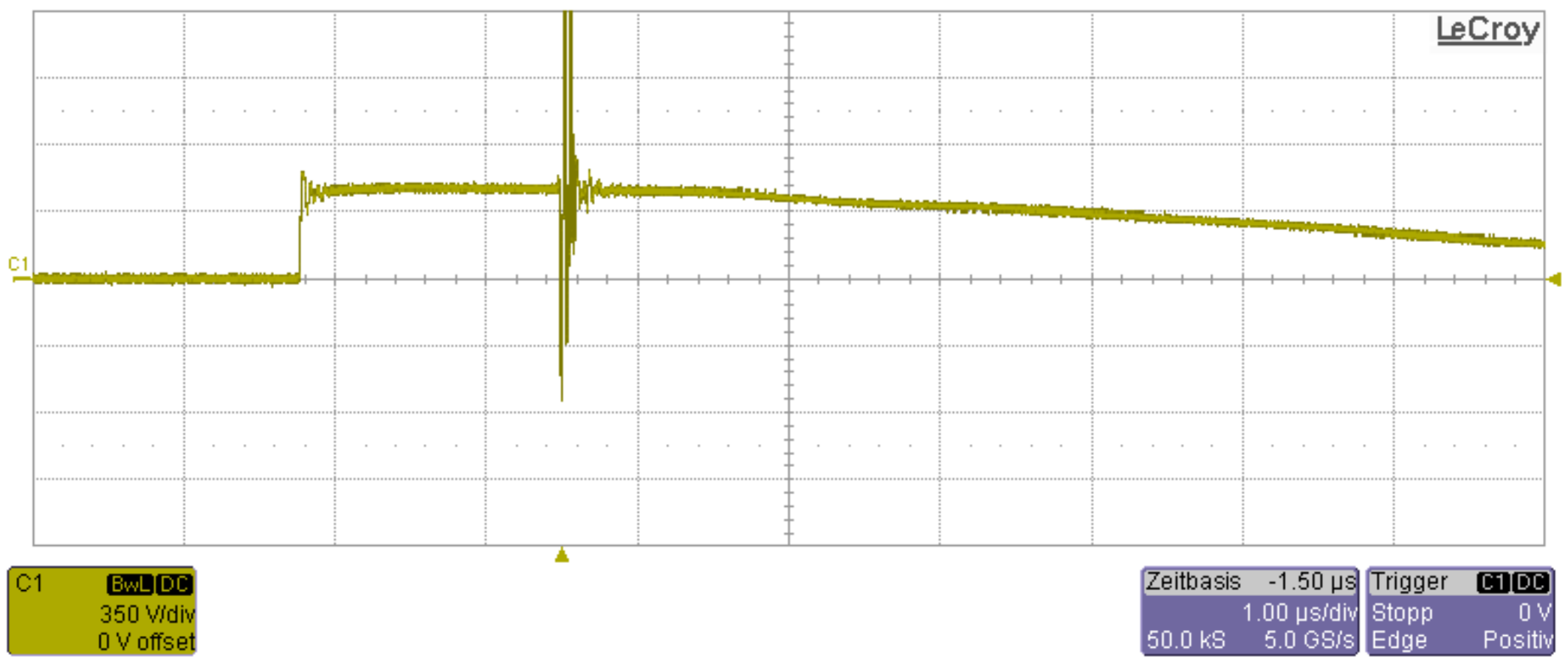

Fig.13.2: V2, ~~BD~~

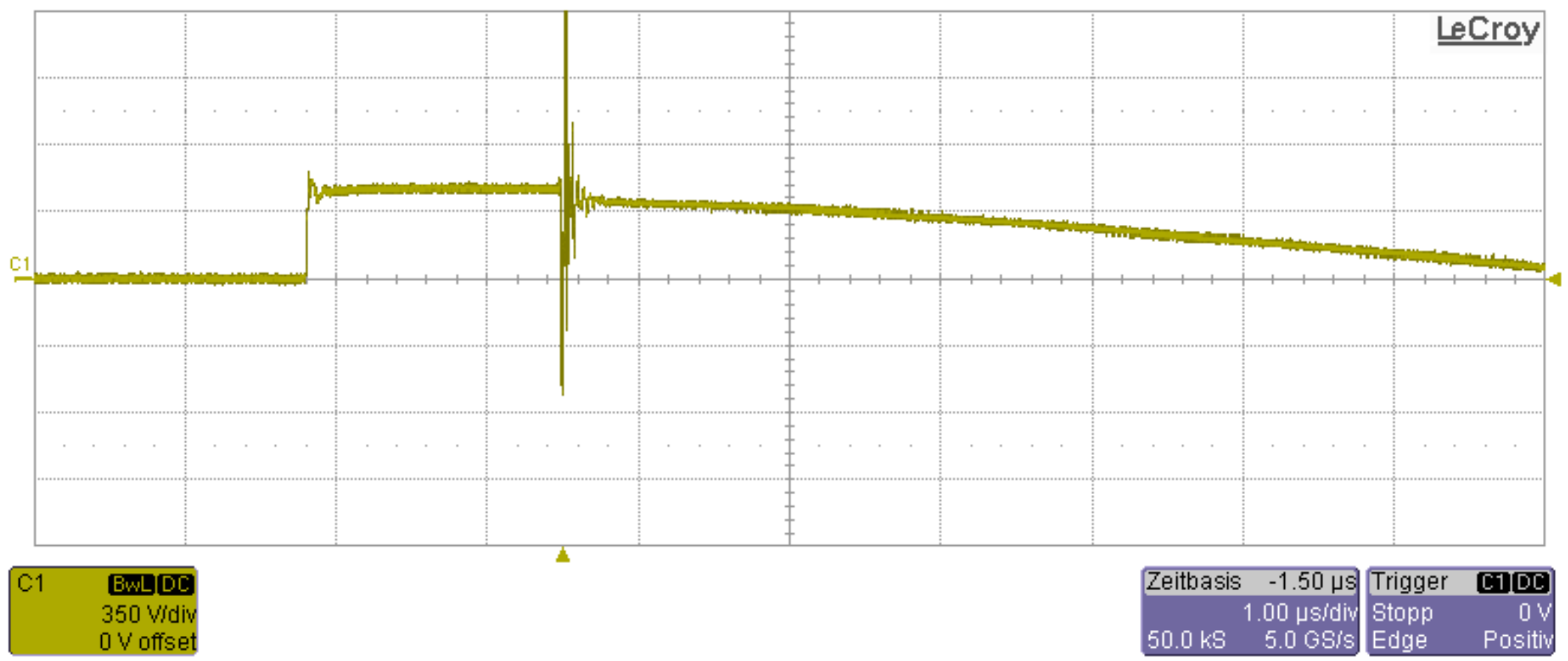

Fig.13.4: V2, BD

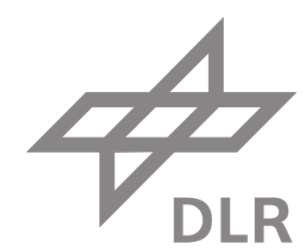

Oscilloscope measurements, 200ns

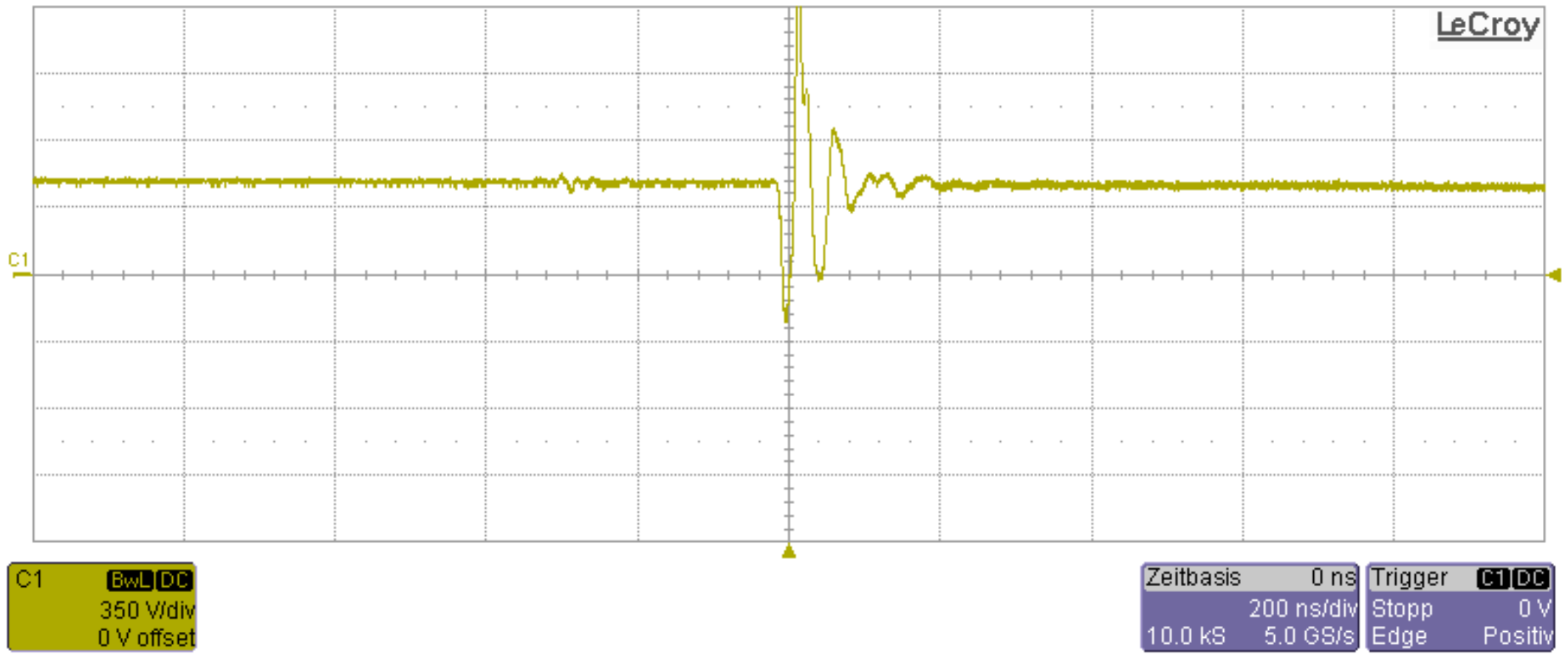

Fig.14.1: V1, ~~BD~~

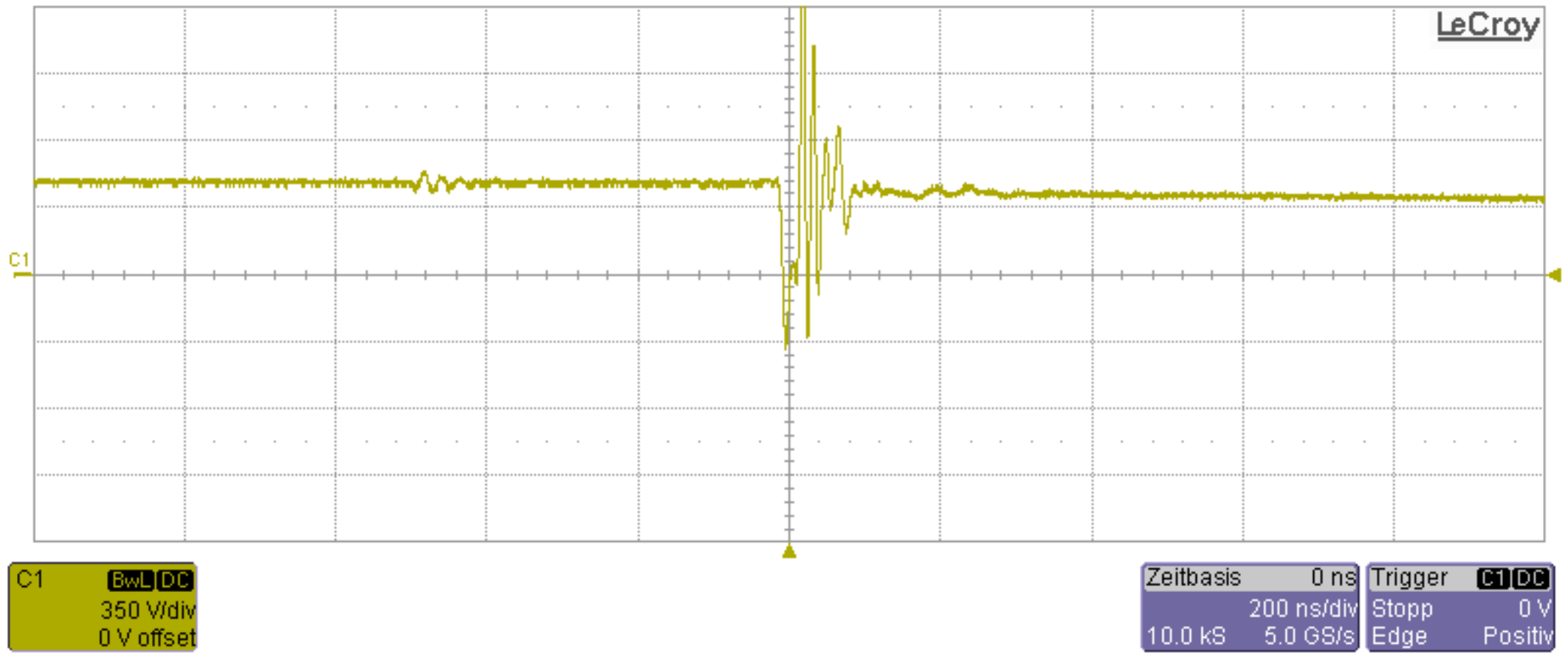

Fig.14.3: V1, BD

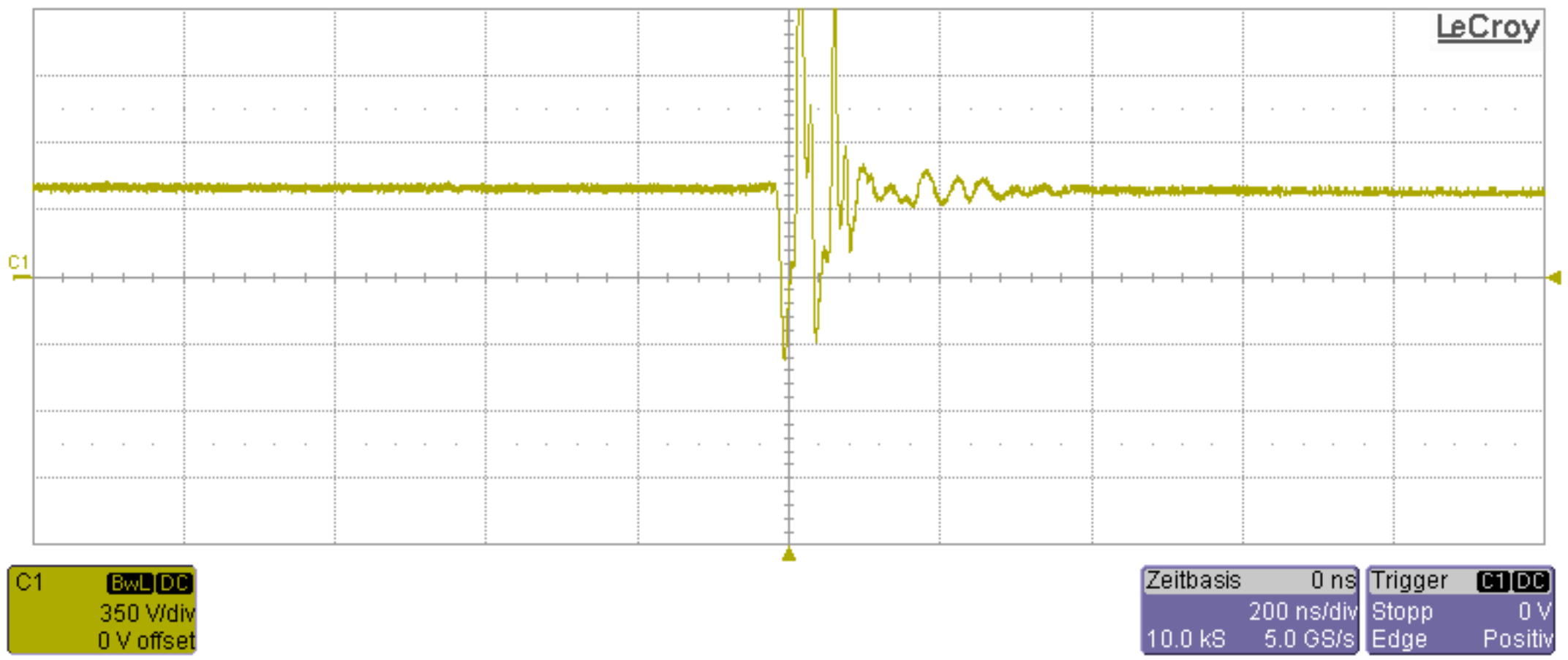

Fig.14.2: V2, ~~BD~~

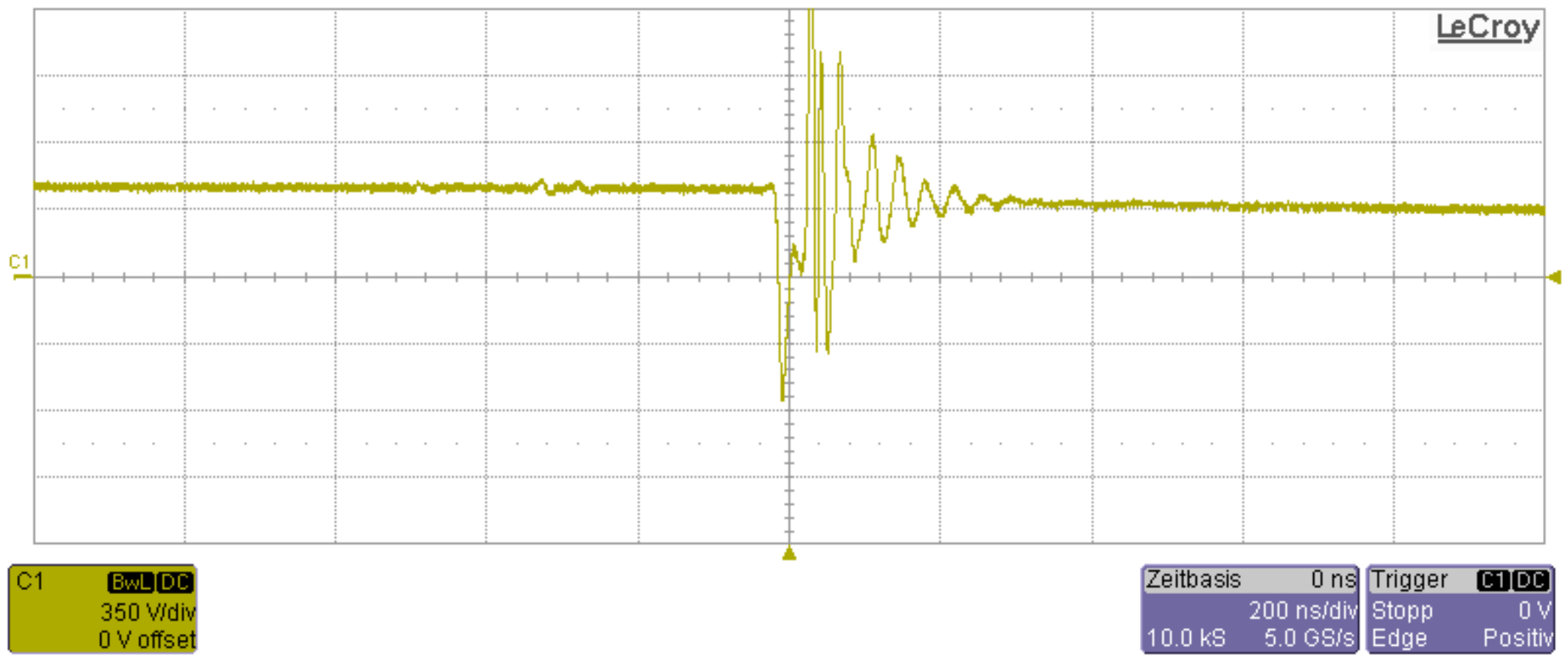

Fig.14.4: V2, BD

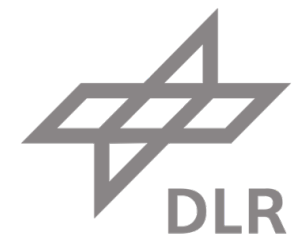

## 4.4 Plasma switch scope measurement analysis

One can notice a lot in common comparing to the figures of hand switch. However following differences can be observed.

T1, switch on time. At the V1 point, the switch on process can be identified by the small perturbation in the voltage, Fig.13.1 and Fig.13.3.

T3, primary coil charge time. For non-BD case this lasts for a ≈1us shorter (≈2us) than with BD case (≈3us), Fig.13.1 and Fig.13.3. Another observation, that 0.5us before the T4, for BD case, there is a voltage perturbation present, Fig.13.3. and Fig.13.4

T4, the high peak oscillation. Like before, it starts with a negative peak. However for BD case, Fig.14.2 and Fig.14.4 indicate a skipped next 1 whole oscillation – after the first negative peak, a barely noticeable positive and then negative bounce follows.

T5, once would again, we note kind of chaotic oscillations for case without BD, and dumped harmonic for BD case.

T6, capacitor discharge. One notes, that this proceeds only until voltage equals to 0V. It comes from the obstacle, that gas discharge tube closes current at that voltage, becoming non-conductor.

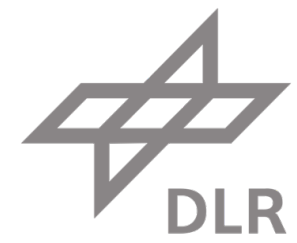

# 5 Photo-based visual analysis of the plasma boom

## 5.1 Air medium photographs

The following is setup within laboratory was built. The red circled part is the photographed object.

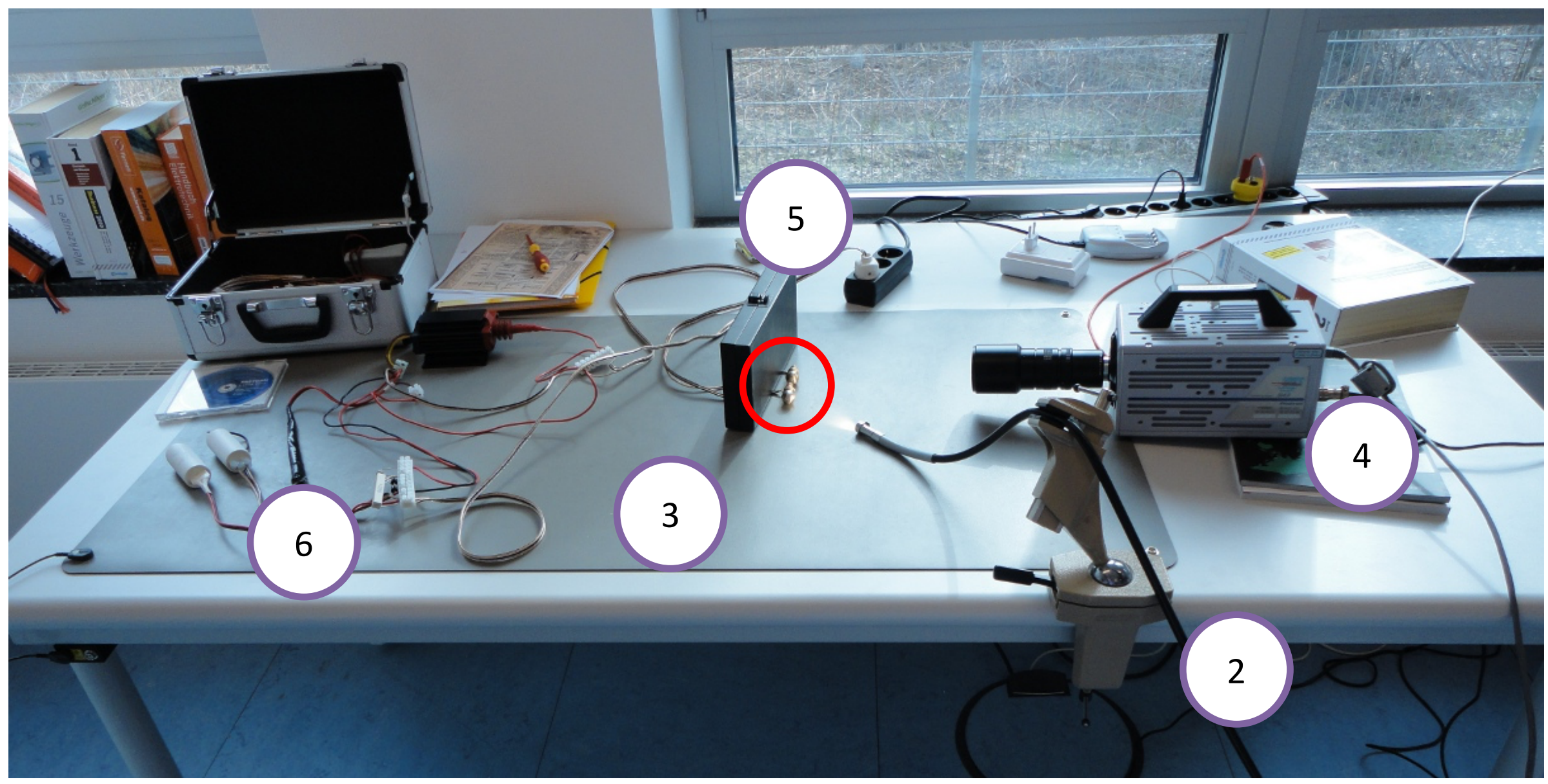

*Fig.15: Plasma spark photographing setup*

1. (red circle) the photographed object
2. The additional light source
3. Grounded table plate?!
4. High speed camera Photron Fastcam SA3, working with 6000 frames per second
5. Black background for photographs
6. The plasma generator circuit, as in Fig.3.

Here is a color photo of the electrodes, made with basic photo camera. For clearer spark observability purpose, one electrode was chosen to be a copper string. Since then it comes very obvious that plasma is extending beyond the electrode itself.

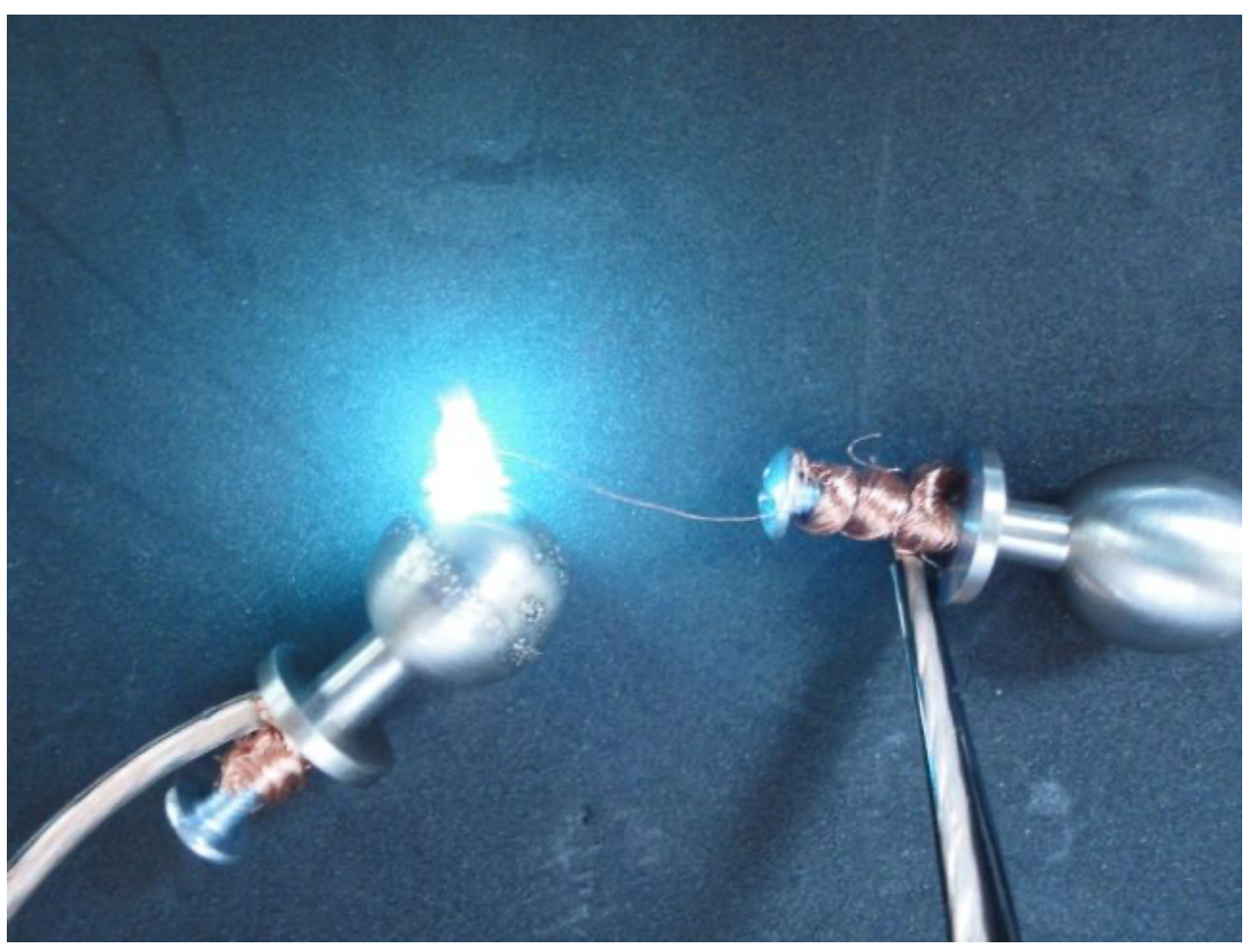

*Fig.16: Color photo of the boosting effect, made with basic camera, look similar to eye observation.*

Fig.17: Three photo sequences of spark discharge, the configuration without booster diodes. The ball is HV positive electrode.

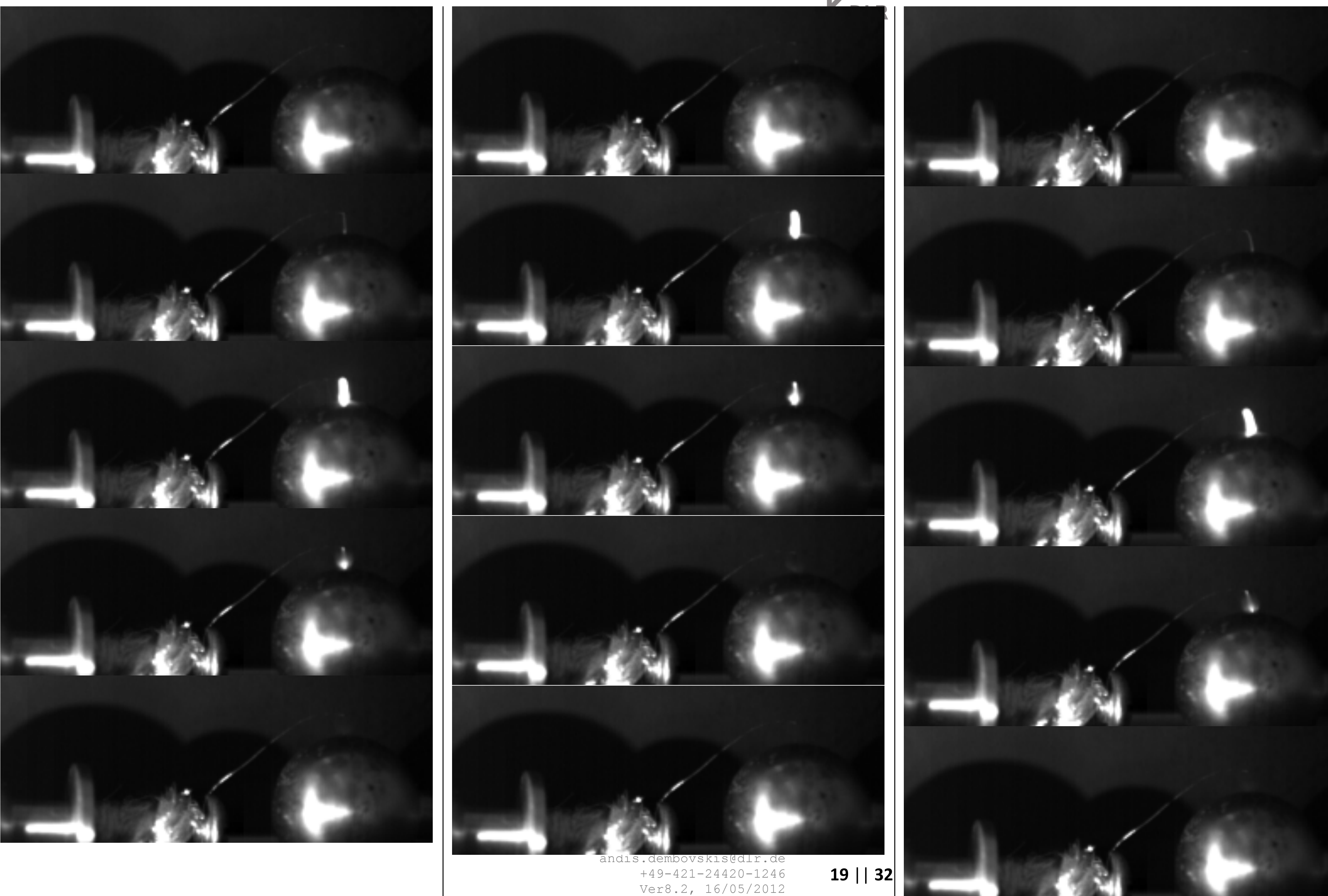

Fig.18: Three photo sequences of spark discharge, the configuration with booster diodes. The ball is HV positive electrode.

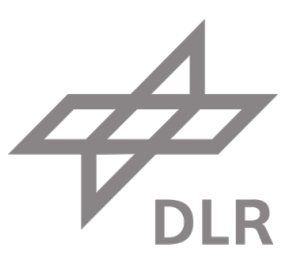

Fig.19: Three photo sequences of spark discharge, the configuration with booster diodes. Configuration changes with respect to Fig.14: 1) The ball is ground. 2) There is a pill of tap water on the top of the metal ball. 3) Externally applied light source has another angle and proximity. 4) The background black metal case is farther away.

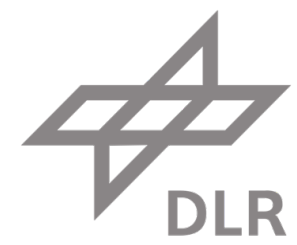

More showcasing photos and of original resolution from the high-speed camera can be found here: http://www.andis.me/pub/plasma_photos.zip

And here are some other showcases of the boosted plasma effect, made with basic color camera.

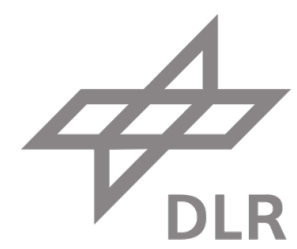

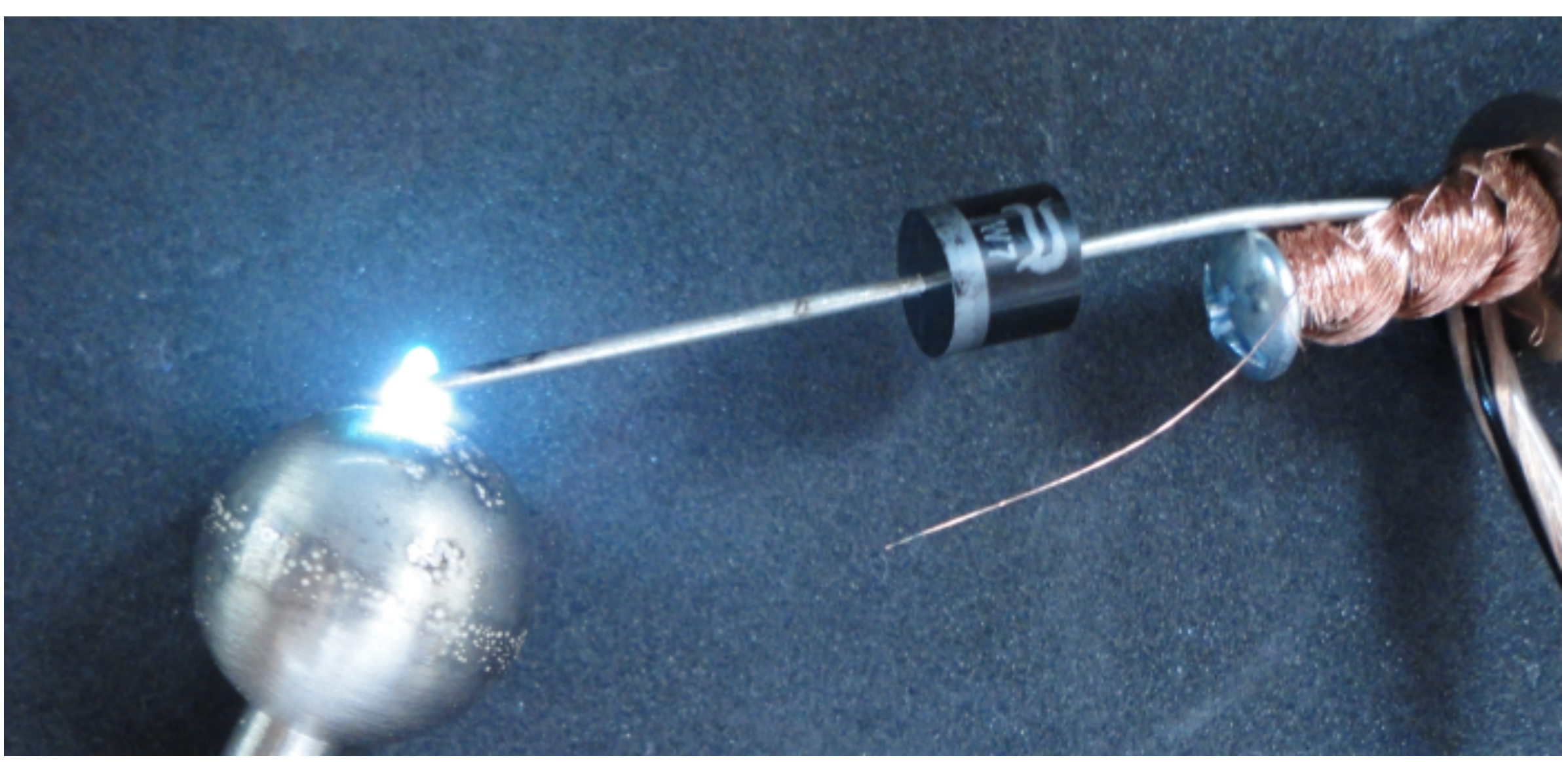

*Fig. 20: plasma pictures with basic colour camera*

## 5.2 Air medium photograph analysis

The following differences one will notice between boosted and non-boosted plasma spark cases:

- With BD, a plasma ball is formed, of round/oval shape, Fig.19.1.3
- With BD, there is no HV path pre-selection. If looking at no BD cases Fig.17.1.2 and Fig.17.3.2, one will notice that HV is creating first a path to go through and then in upcoming pictures the path is ignited as spark. Whereas in the case with BD, the Fig.21.1, doesn't show such a path-creation before high plasma arise.

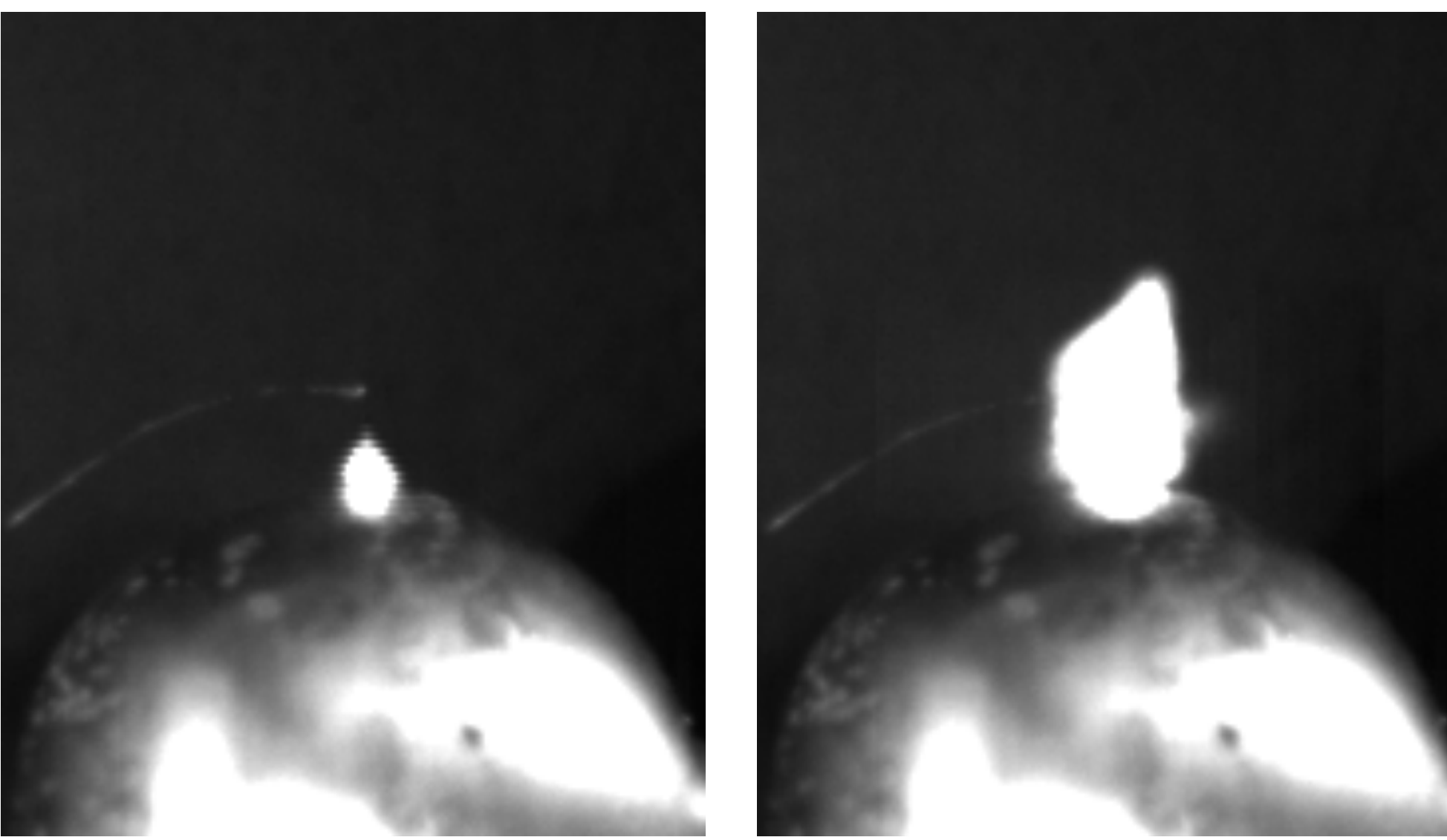

*Fig. 21.1 and 21.2: original resolution of photos Fig.18.2.1 and Fig.18.1.2 respectively.*

- With BD, the plasma is flowing out of source electrode like "gas" toward target electrode, Fig.18.2.1. And like a "gas flow", it does with its inertia pass over the target electrode, Fig.18.1.2, Fig.18.3.1, Fig.19.1.2, Fig.19.2.3, Fig.19.3.2.

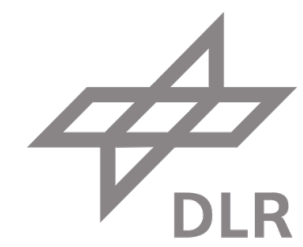

- With BD, the visible plasma lasts for 4 frames (4x 1/6000[s]), whereas without BD, plasma lasts only for maximum 3 frames.
- With BD, plasma is starting its forming at HV positive electrode, the Fig.18.2.1, Fig.19.1.2.
- Plasma finishing: without BD it ends up forming at positive HV electrode: Fig.17.1.4 Fig.17.2.3 Fig.17.4. With BD, it is somewhere between the electrodes Fig.18.2.3 and Fig.18.2.4 or even away: Fig.19.3.4 and Fig.19.3.5.
- With BD, water spill on the electrode didn't have a noticeable influence to the booster effect. The same for case without BD (not showcased here in photos).
- Observing the booster effect with eye, it looks similar to Fig.12.
- Observing the booster effect with eye, the color of it seems white. The color of discharge without BD, looks blue.
- With BD, there is a loud bang when discharge occurs (audibility measurements should be performed). Like air pressure shockwave.
- Within hand switch configuration(Fig.4.), with 3mm gap between discharge electrodes, when turning source voltage down: without BD the spark flash don't exist below 160VAC, whereas with BD, the spark is produced until 70VAC. This shows, that with BD configuration, the flash-over has higher energy level.
- With BD, in photograph sequence there could be observed that the copper string is swinging after the bang. The first move is "away from the other electrode". The string oscillations last for approximately 2 seconds (almost zero at beginning of the next bang). The electrode string oscillations can be also observed by eye as "smudged string". Oscillation frequency is approximately 60 frames (10[ms]) at approximately 1mm away from center.
- It was noticed, that if one of electrodes is a ball-type, the discharge-effect is bigger louder.
- The effect is present with HV back diode (Fig.20.3). Meaning, discharging from HV electrode towards diode, which don't let positive charge to flow towards negative side, also persists. Diode-electrode can be positioned both directions, not alternating the plasma effect.
- The boosting effect is not a DC current discharge. Since, considering Fig.5.3, one sees that the capacitor stays charged after the spark. The same can be concluded from Fig.12.3 – as soon as voltage drops to 0V due LC oscillation in capacitor->plasma_switch->primary_coil circuit, the capacitor is switched away from spark circuit already at the middle of T6, whereas spark occurs and becomes conductive only starting at T7 for 10us (Fig.8.2).

Conclusion1: the BD setup generates plasma ball instead of plasma string.

Conclusion2: with BD the discharge is longer

Conclusion3: with BD the discharge is of bigger volume

Conclusion4: with BD the discharge is accompanied by "strong" audible bang

Conclusion5: with BD the color of discharge is white

Conslusion6: it is not clear from classic electrodynamics perspective, why the BD effect persists

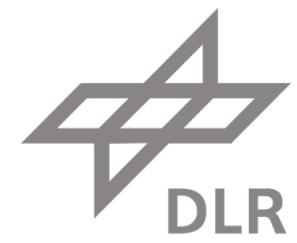

## 5.3 Photographs in water medium

Here are two photographs, a steady frame cut out from video, made with basic color photo camera of the spark within water (0.5l tap water, hot, with 3 tablespoons of salt dissolved).

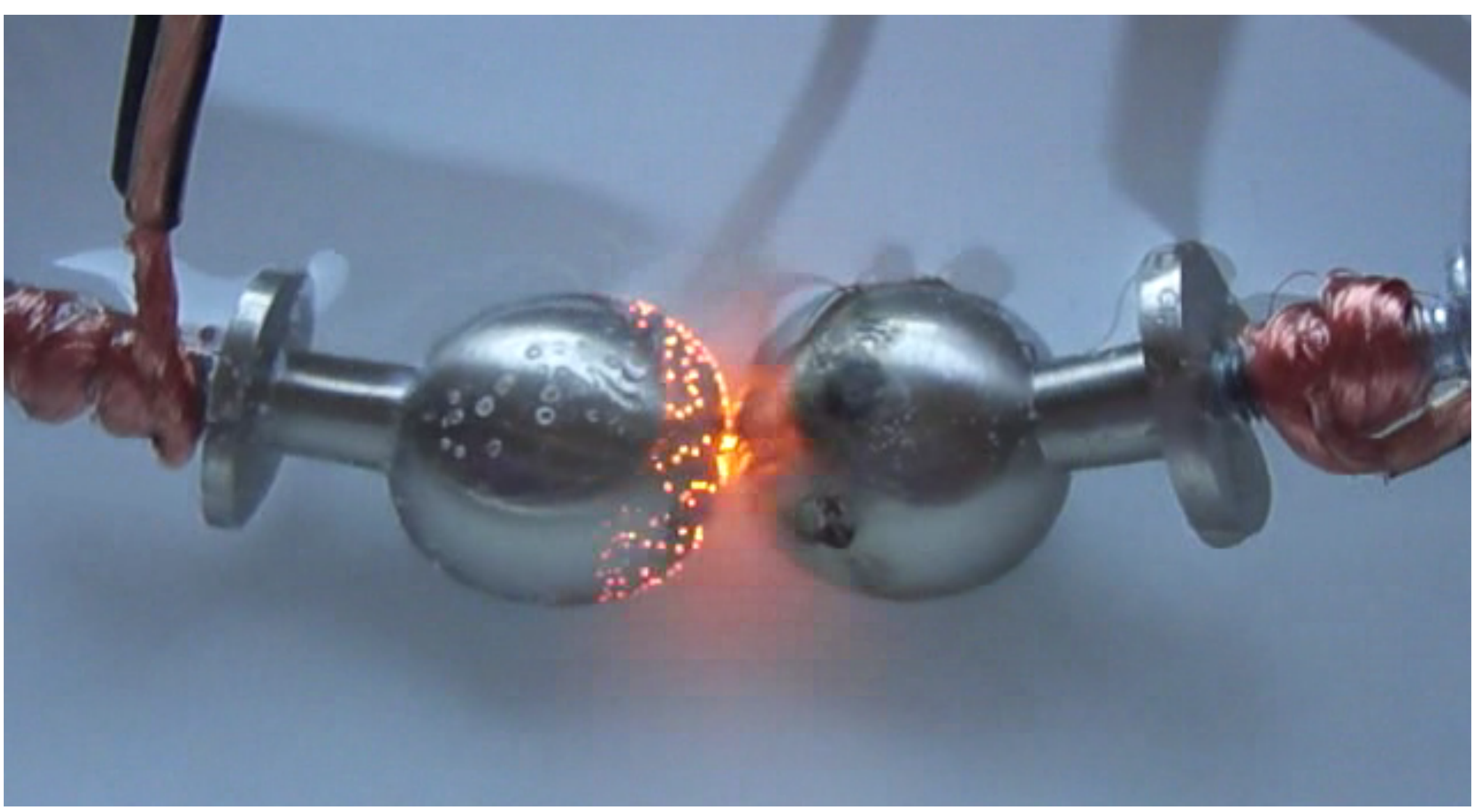

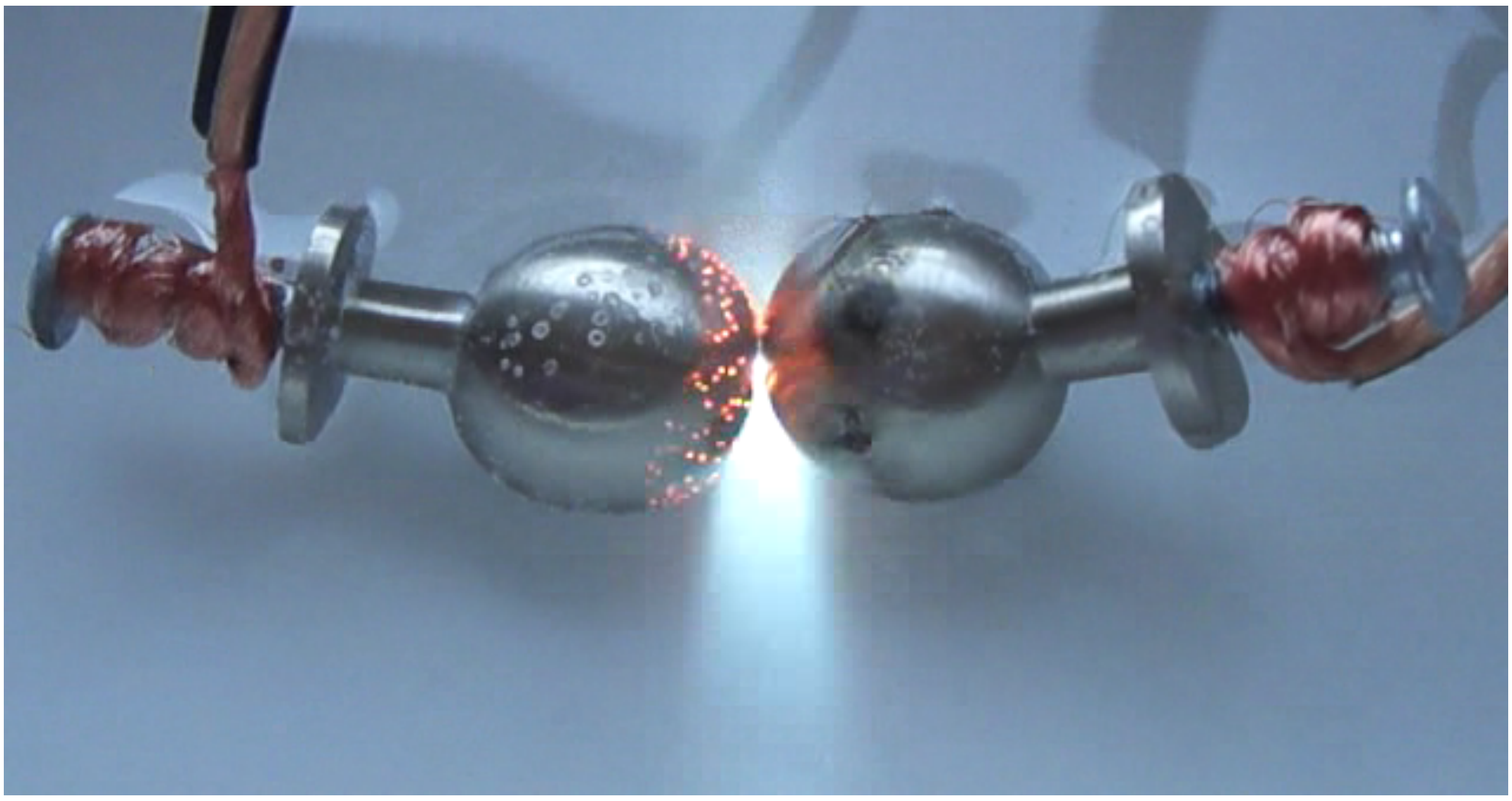

*Fig.22: Electrodes immersed in salt water*

The left electrode is the negative, the right is the positive HV.

The spark was present only if electrodes were very near, only in water with electrolyte. There were visible red light points on top of the square around the negative electrode. The bang from discharge ruffled water and sound like balls hitting the bottom of the water dish. It was not clear whether the bang is a shock wave from plasma discharge or the ball-bouncing because of repulsion power. It sound like metal ball bouncing against the bottom of the dish.

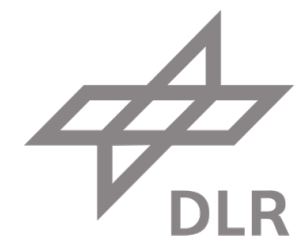

A feasible explanation of this could be that there was a hydrogen generated with the discharge (this was obvious) due classic electrolysis, and that the HV did ignite it within the water.

It would be of interest to check discharge event in the case of evacuated / partially evacuated air environment.

# 6 Preliminary modeling analysis

From classical electromagnetism theory, we know that EMF can be calculated by formula:

$$\varepsilon = -N \frac{d\phi_B}{dt}$$

where

- ε - electromotive force
- N – number of turns in windings of transformer
- ϕ – magnetic flux

For simplicity let us assume $V_i$ to be input voltage, $V_o$ to be output voltage, than we have relation

$V_o = -N \frac{dV_i}{dt}$ , for sinusoidal wave giving us the following phase delayed response signal as in Fig.23.1 below.

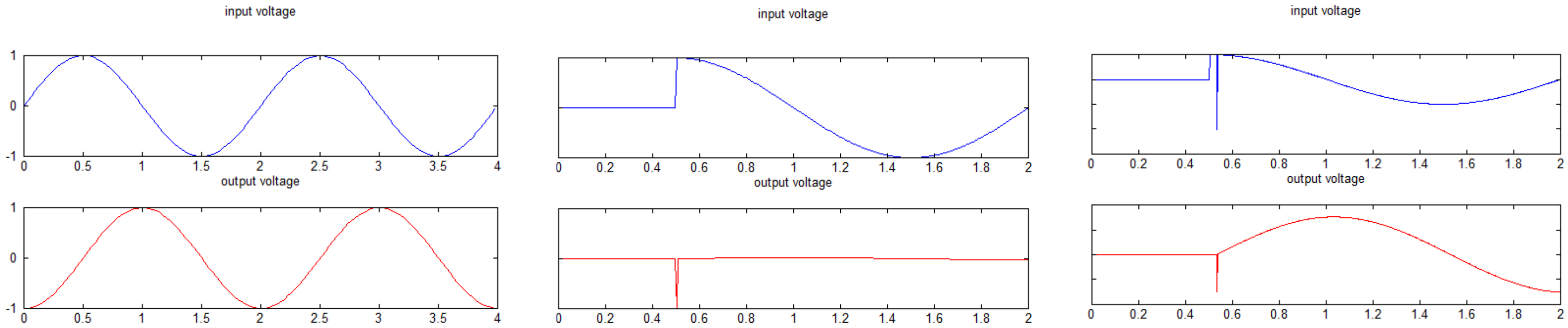

*Fig.23. Transformer response to input voltage: (1) 90⁰ delay (2) step function to Dirac function (3) Dirac function with tiny delay added to input function through BD circuit*

Considering that we have switch-capacitor-inductor cycle (the left part of Fig.2a), it is correct to model the switch on time as step function with supplied voltage to the coil.

The derivative of step up function is Dirac function and since we have the minus sign in front, we get the following conversion:

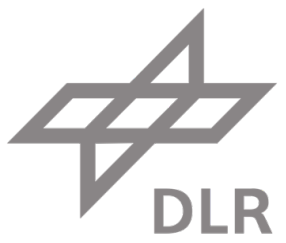

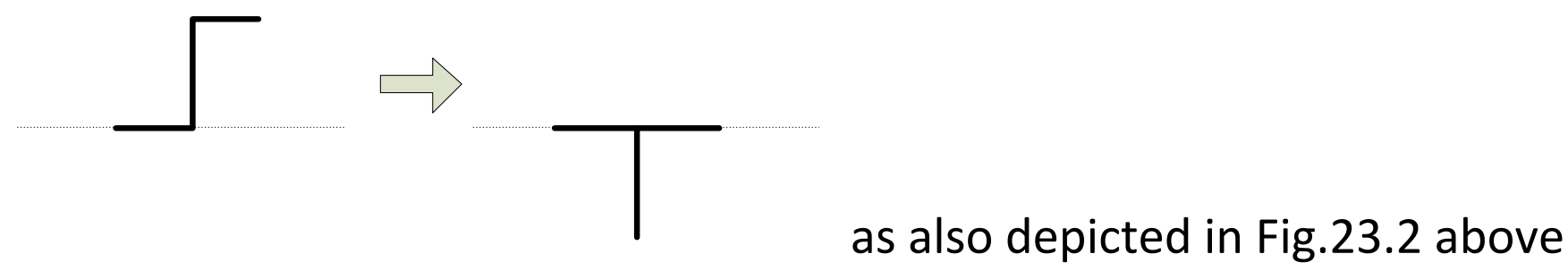

as also depicted in Fig.23.2 above.

And since the Booster-Diodes, as showed in Fig.2b are open for negative voltage flow towards positive electrode, the switch-on action induced negative High Voltage after induction-used time delay (T2) could go this path and pose a deep negative voltage drop to the inputs of first coil (T3), Fig.23.3. Since this sudden voltage change magnifies oscillation amplitude at first coil, this could manifest as higher voltage output at second coil. The same amount of charge, but condensed to higher voltage, thus being capable to initiate plasma discharge at higher energy.

However this explanation is not supported by following two observations:

- Additional forward diodes at coil output, before dividing to cable of diodes and discharge gap, didn't influence the effect – meaning negative current couldn't flow through it to capacitor.
- Dirac down-pulse at input after switch-on, within the T3 does occur also when operating without BD. Also it did not reliably showed to be more intense in BD case.

Further analyses are needed, to check if this theory holds true, since negative pulse at T3 was measured also for non-BD case (see Fig. 6). An analysis on when does the plasma discharge occurs compared to voltage measurements as well as output voltage measurements could greatly contribute to understand further details on this experiment.

# 7 Spectrum of plasma discharge

Here are few preliminary plasma spectrum records, measured with Ocean Optics HR2000+CG-UV-NIR. Thanks for support to University of Kiel, Institut für Experimentelle und Angewandte Physik, AG Plasmatechnologie.

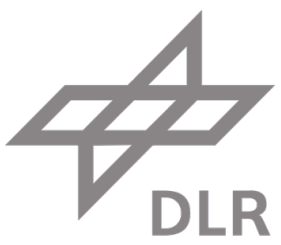

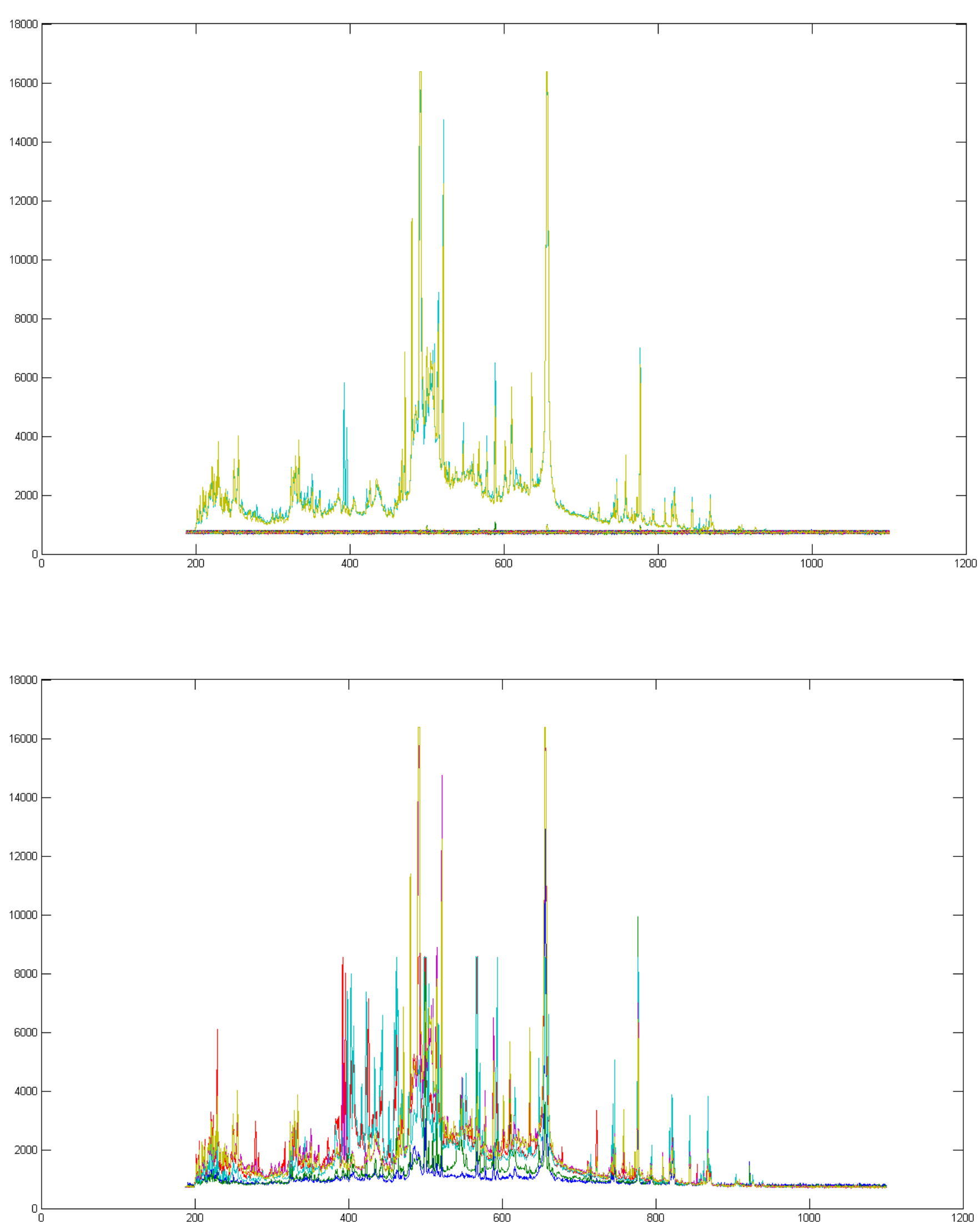

Fig.24: Plasma discharge spectrum. (1)two plasma discharges, within continuous sampling mode (2)overlapped 6 plasma discharges, recorded by different angles and proximities.

# 8 Choice of setup parts

Electronic parts were chosen with following arguments:

- Capacitors: the motor starting capacitors were chosen, as they offer good figures for speed of current release. Also non-polar capacitor choice was made because of having fluctuations in circuit. From Conrad.
- Diodes: as diodes everywhere were chosen 1N5408, for price and HV withstanding reason. At the BD package, 6..10 diodes were used, since one would burn out with first try. There one could have chosen using instead 1x very high quality HV diode. From Conrad.

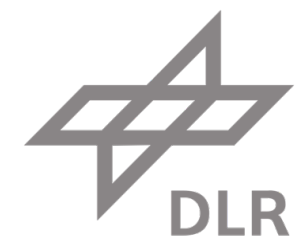

- DPDT switch: was chosen with rating 220V, 30A. This was the best in Conrad and sufficient for the application.
- Gas discharge tube: the LittleFuse CG2 series product was used due good properties of surge current levels, peak current levels, max voltage levels, good lifetime characteristics. From Farnell.
- Electrodes: copper string was taken from the simplest electricity cable. The metallic ball was of random choice. The spark might differ for case with copper-copper electrodes, since having more free electrons and better conductivity. From Conrad / Household market.
- The ignition coil: required is CDI coil, was chosen CraneCrams PS91 with 1:54 turns ratio, 5.5mH inductance, 0.43Ohms primary resistance [4]. A better choice might have been with 1:100 ratio. Was imported from USA, eBay. At later stages coils used were Accel 8140C with 1:100 turns ratio and Accel 140019 with turns ratio 1:94 (depicted in Fig.25.).

The overall cost of the setup parts lies around 250EUR.

# 9 Setup and test quality considerations

There might molt of quality and especially security improvements, by substituting parts with other ones of higher quality, including the basic 220V rated cables with HV cables.

One week point for the voltage measurements was non-deterministic wire alignment on the table and no magnetic isolation to wires, allowing cable interactions via magnetic fields. This could be partially solved by using shielded cables.

Also it would have been of great value to have higher range HV probes for oscilloscope, also some options for measuring current as well as sound and light and heat intensity effects for comparison.

In one of experiments [1] was noted that the spark with BD didn't heat the electrodes, whereas classic spark did. This is much unexpected and would be high interest to be verified.

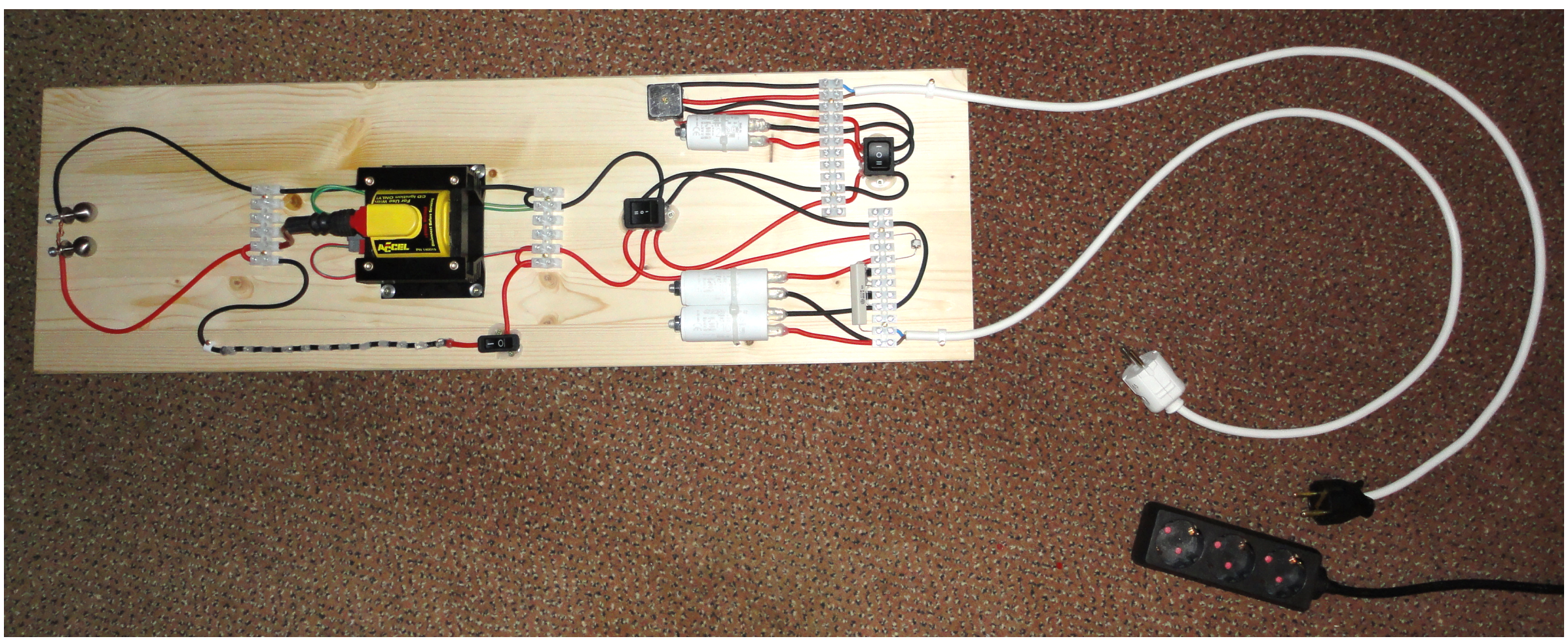

*Fig.25: A close look to setup as in Fig.3*

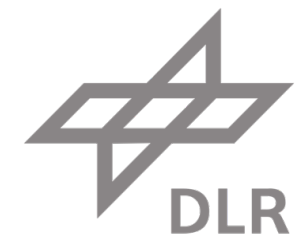

## 10 Proposed explanation of the effect occurrence

From classical physics and electronics point of view it is not clear why the effect is observed. However there are several ideas proposed, none jet been approved. Few of them:

a) Initially the generated HV ionizes spark gap (although not seen in photos) and that shortens the capacitor discharge over spark, making it very fast (although in oscilloscope measurements discharge speed is barely higher; and within photographs the plasma visibility time is longer in the case with booster-diodes).
b) Resonance of potentials: the fast voltage on-switch time induces a "small" negative HV at secondary, which flows back through BD diodes introducing additional HV drop at primary (although not seen in oscilloscope measurements), resulting to boosted HV at the output of the secondary winding.
c) Current flow acts like incompressible gas inertial movement: electrons which participate in current movement have "momentum" – like flow of water. As soon as they meet an abrupt obstacle in one path, they would compress against and generate a reward flowing wave with higher potential. [1, p97]
d) Aether[1] contributes to flashing: it is proposed by Nikola Tesla, that a flowing current consists of moving electrons and aether combined. Thus could be interpreted following versions:
    a. initially current flows against diodes, when abrupt obstacle met, the electrons, abruptly stop, whereas aether acts like compressible gas – "stretching" and "bouncing" back, flowing along the cable toward discharge gap and contributing to the HV discharge giving the boosting effect.
    b. when the abrupt obstacle met, the aether spreads out as "radiant energy" [2], contributing to conductivity/ionization of the air in spark gap.

## 11 Practical applications

The following properties of the boosted spark are expected to play improvements in internal combustion engines for cars as well as turbine ignition systems for airplanes:

- The plasma intensity is much higher – this can ensure more robust fuel ignition, also for cold engine
- The ignition front is bigger – facilitating pickup of burning front. This allows more rapid fuel combustion within engine, leading to more efficient ignition timing option
- The bang expansion – the spark is accompanied with rapid pressure expansion wave – facilitates speed of burning front expansion
- Water moisture do not disturb – moist addition within air of fuel mixture will not disturb explosion of the spark (could significantly help for proper and more robust turbine functioning in rainy weather)

The following pictures sketch approach differences for classic CDI installation and proposed CDI ignition with boosted plasma (CDIwBP) for internal combustion engines:

[1] Another substance than the aether as defined in Michelson-Morley experiment, prooved to non exist.

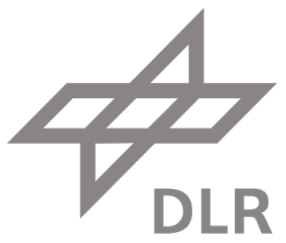

Basic CDI discharge in cars

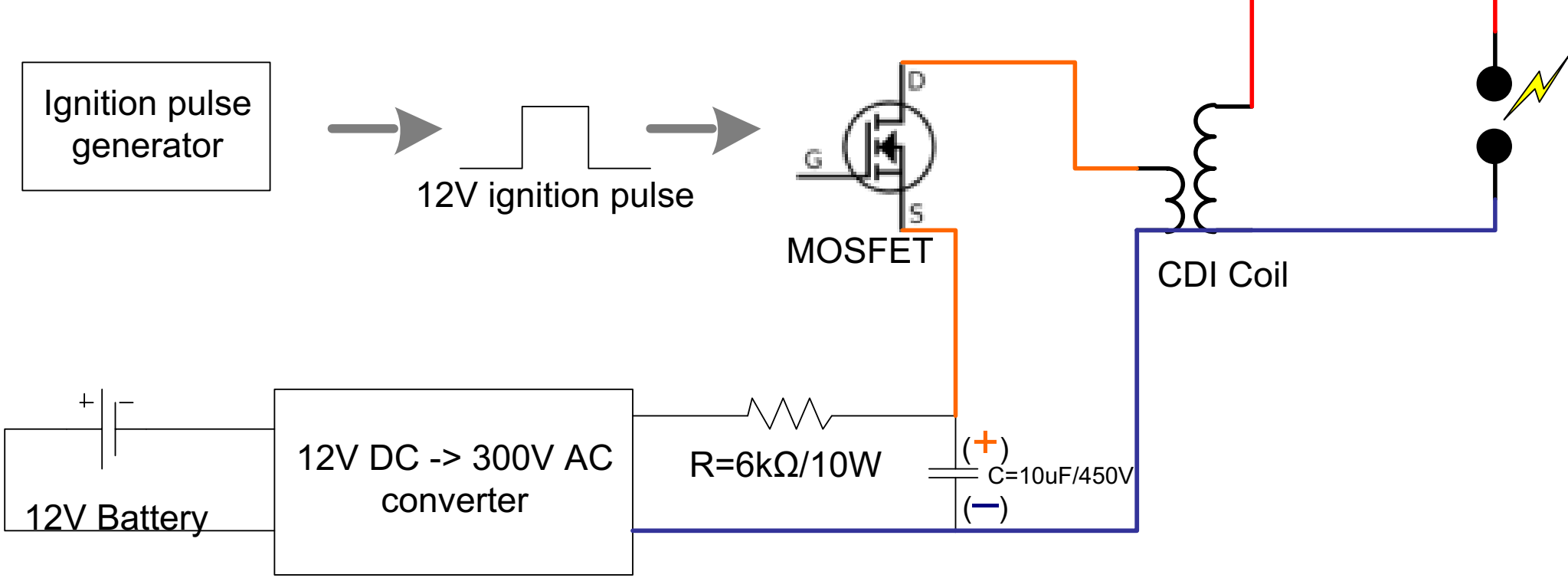

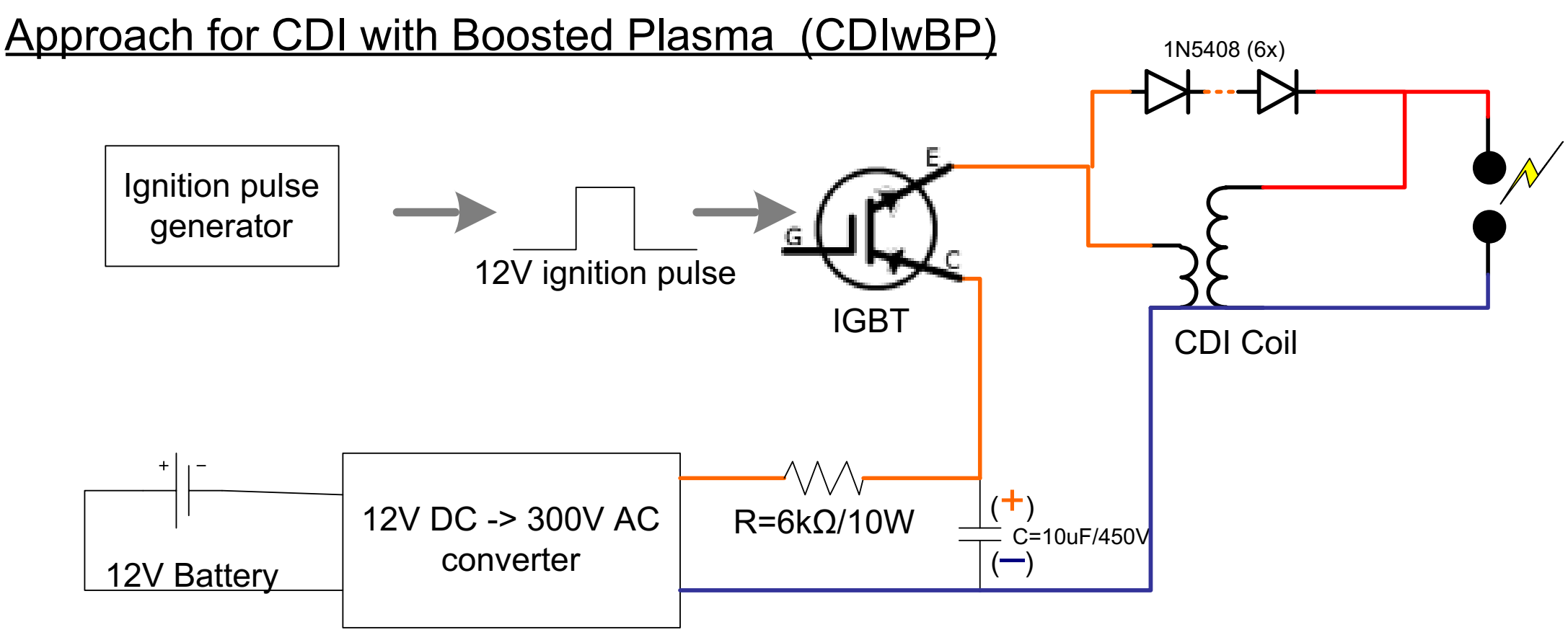

*Fig.26 Classic CDI and boosted plasma CDI circuit principal scetches*

There are two principal differneces between the schematics:

1. In CDIwBP there is IGBT switch instead of MOSFET used – this is needed because of requirement for rapid switching of time near to 20ns.
2. In CDIwBP there is an additional diode circuit installed, to give the boosting effect, described in this paper.

Following optimizations would help to further decrease capacitor charging power:

1. Tuning IGBT ignition pulse length to exactly match switch off at T6, before T7, thus saving unnecessary voltage potential in capacitor with opposite sign.
2. Additional circuit to swap polarity of capacitor after each ignition occurrence – so that charging capacitor would use just approximately 1/4$^{th}$ of its power capacity.

# 12 Outlook

An additional evaluable analysis [2] could be made by observing the discharge effect by driving alike test, whereas disconnecting capacitor from primary coil circuit (containing the BD) at T3, T5 or the beginning of T6. It would help to indicate the significant timing for voltage application to manifest the HV and the boosting output. To carry this out, one would need high end IGBT transistors, programmable signal generator up to 1GHz, also an oscilloscope with 1GHz probes and at least 2kV range.

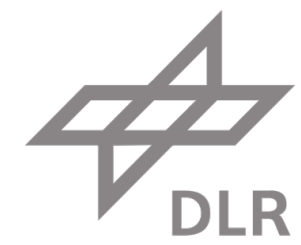

There could be mathematically made a circuit resonance simulation, to check if that could be source of boosting effect, although there are no voltage changes observed in the input characteristics at primary coil input with or without BD. To match the simulation results with oscilloscope measurements, one would need a very precise coil properties measuring toolset. It would be of high interest to do this circuit simulation by using equations from Weber's Electrodynamics [8] instead of Maxwell Electrodynamics.

Although further research for an improvement of ignition system applications could be carried out, the current state of the observations and results allow implementing significant improvements for internal combustion and turbine engines.

# 13 Notes

- It would be of very high practical value to test this setup on a regular motor or a car. In an unofficial literature one can read that a big plasma discharge ball increases power of the car's motor leading to 15% fuel economy as well as emission drop by 1/3 when combined with water mist injection [1],[3],[5].
- It would be very useful to have HV measurement equipment available – that would allow better inspection of current and voltage flows within the circuit.
- As a better alternative switching element, a distributor from car ignition system could be used to charge and discharge the capacitors.
- It would be of high value for research to replicate Nicola Tesla's device of disruptive HV DC current circuits to observe and test effects of aether flow as described by him.